\documentclass[fleqn,usenatbib]{mnras}

\usepackage[T1]{fontenc}
\usepackage{ae,aecompl}

\usepackage{amsmath}	
\usepackage{fontawesome}
\usepackage{enumerate}
\usepackage{algorithm}
\usepackage{algorithmicx}
\usepackage{algpseudocode}

\usepackage[hang,singlelinecheck=false]{caption} 
\usepackage{subcaption}
\usepackage{graphicx}
\usepackage{multicol}
\usepackage[normalem]{ulem}
\usepackage{tabularx}
\usepackage{threeparttable}
\usepackage{xcolor}

\usepackage{newtxtext,newtxmath}

\newcommand{\numlens}{23}

\newcommand{\editref}[1]{{{#1}}}
\newcommand{\editreft}[1]{{{{#1}}}}
\newcommand{\editrefth}[1]{{{{#1}}}}

\newcommand{\editf}[1]{{#1}}
\newcommand{\edits}[1]{{#1}}

\newcommand{\editth}[1]{{#1}}
\newcommand{\editfr}[1]{{#1}}
\newcommand{\ttedit}[1]{{#1}}
\newcommand{\editfv}[1]{{#1}}
\newcommand{\editsx}[1]{{#1}}
\newcommand{\editsv}[1]{{#1}}
\newcommand{\editet}[1]{{#1}}

\title[Elliptical galaxy structure]{Dark matter halos of massive elliptical galaxies at $z\sim 0.2$ are well described by the Navarro--Frenk--White profile}

\author[Shajib et al.]{
Anowar J. Shajib$^{1, 2}$\thanks{E-mail: ajshajib@uchicago.edu},
Tommaso Treu$^{2}$\textsuperscript{\thanks{Packard Fellow}},
Simon Birrer$^{3}$ and
Alessandro Sonnenfeld$^{4}$
\\
$^{1}$Department  of  Astronomy  \&  Astrophysics,  University  of Chicago, Chicago, IL 606374, USA \\
$^{2}$Department of Physics and Astronomy, University of California, Los Angeles, CA 90095, USA \\
$^{3}$Kavli Institute for Particle Astrophysics and Cosmology and Department of Physics, Stanford University, Stanford, CA 94305, USA \\
$^{4}$Leiden Observatory, Leiden University, Niels Bohrweg 2, 2333 CA Leiden, the Netherlands
}

\date{Accepted XXX. Received YYY; in original form ZZZ}

\pubyear{2020}

\begin{document}
\label{firstpage}
\pagerange{\pageref{firstpage}--\pageref{lastpage}}
\maketitle

\begin{abstract}

We investigate the internal structure of elliptical galaxies at $z\sim 0.2$ from a joint lensing--dynamics analysis. We model \textit{Hubble Space Telescope} images of a sample of 23 galaxy--galaxy lenses selected from the Sloan Lens ACS (SLACS) survey. Whereas the original SLACS analysis estimated the logarithmic slopes by combining the kinematics with the imaging data, we estimate the logarithmic slopes only from the imaging data. We find that the distribution of the lensing-only logarithmic slopes has a median $2.08\pm0.03$ and intrinsic scatter $0.13 \pm 0.02$, consistent with the original SLACS analysis. We combine the lensing constraints with the stellar kinematics \editref{and weak lensing measurements,} and constrain the amount of adiabatic contraction in the dark matter (DM) halos. We find that the DM halos are well described by a standard Navarro--Frenk--White halo with no contraction on average for both of a constant stellar mass-to-light ratio ($M/L$) model and a stellar $M/L$ gradient model. For the $M/L$ gradient model, we find that most galaxies are consistent with no $M/L$ gradient. Comparison of our inferred stellar masses with those obtained from the stellar population synthesis method supports a heavy initial mass function (IMF) such as the Salpeter IMF. We discuss our results in the context of previous observations and simulations, and argue that our result is consistent with a scenario in which active galactic nucleus feedback counteracts the baryonic-cooling-driven contraction in the DM halos.
\end{abstract}




\section{Introduction} \label{sec:intro}


\editfv{Measurements of} structural properties of elliptical galaxies can test prediction of galaxy formation theories in the cold dark matter (CDM) paradigm \citep[e.g.,][]{Dubinski94, Kazantzidis04, Debattista08, Read14}. In  this paradigm, small haloes merge to hierarchically form larger halos. CDM-only $N$-body simulations predict that the dark matter is universally distributed according to the Navarro--Frenk--White (NFW) profile with a `cuspy' central slope, \editf{i.e., the 3D density scales as $\rho \propto r^{-1}$ in the inner region} \citep{Navarro96, Navarro97}. \editf{Such slopes have been observed in galaxy clusters \citep[e.g.,][]{Limousin07, Caminha17}. However, shallower central density slopes also have been observed in some galaxy clusters and in dwarf and low-surface-brightness galaxies \citep[e.g.,][]{deBlok01, Sand08, Oh11, Newman13}. \edits{One possible explanation of these shallower slopes is given by alternative dark matter models,} e.g., self-interacting dark matter and warm dark matter \citep[e.g.,][]{Dodelson94, Spergel00, Colin00}.} \edits{Another possible explanation, instead, is related to the astrophysics of galaxy formation and evolution. In fact, as \editth{massive elliptical} galaxies are believed to be the end-product of the hierarchical merging and accretion processes, their mass density profiles are a sensitive probe of the physics of galaxy formation and evolution.}

In the galaxy formation process, baryons play an important role that can affect the central density slope of the dark matter distribution. If gas cools and inflows slowly towards the centre, then the dark matter distribution can be adiabatically contracted \citep{Blumenthal86}. In contrast, the dark matter distribution can expand in dissipationless mergers \edits{or} due to gas outflows driven by supernova feedback, \edits{stellar feedback,} or active galactic nucleus (AGN) feedback \citep[e.g.,][]{ElZant01, Nipoti04, Peirani08, Pontzen12}. \editf{Observational avenues to study these processes and their importance in galaxy formation have been limited. The combination of lensing and dynamics has been one of the most informative probes of the dark and luminous matter distributions in the inner region of galaxies and clusters \editth{\citep[e.g.,][]{Treu02b, Czoske08, Barnabe11}}.} The elliptical galaxies in the Sloan Lens ACS (SLACS) survey were found to have no contraction on average \citep{Dutton14, Newman15}, although individual galaxies can have contracted (or expanded) dark matter distributions \citep[e.g.,][]{Sonnenfeld12}. As many or all of the above-mentioned baryonic processes happen at various points of the galaxy formation process, the degree of contraction or expansion depends on the relative importance of these baryonic processes.

The interplay between baryon and dark matter in galaxy formation is also highlighted by the so-called `bulge-halo conspiracy' \editfr{\citep{Treu06, Humphrey10, Cappellari16}}. This conspiracy refers to the nearly isothermal profiles of the total matter distribution with small scatter ($\sim$0.1--0.2) observed \editf{within half of the half-light radii to 100 half-light radii of elliptical galaxies} \editfr{(e.g., from strong and weak lensing: \citealt{Treu04, Gavazzi07, Auger10b, Ritondale19}; from stellar dynamics: \citealt{Thomas07, Tortora14, Bellstedt18})}. 
As neither of the baryonic and dark matter distributions follows a power law, fine-tuning between these two distributions is required to produce the isothermal distribution for the total mass. To understand the origin of this conspiracy through simulation, we can use two \editfr{parameters}: the dark matter fraction $f_{\rm dm}$ within the inner region and the distribution of logarithmic slope $\gamma$ for the total mass profile. Near the \editf{half-light radius}, dark matter has as a logarithmic slope of $\gamma \leq 1.5$ and the baryonic distribution has a logarithmic slope of $\gamma \sim 2.3$ for a de Vaucouleurs profile \citep{deVaucouleurs48}. Thus, fine-tuning in $f_{\rm dm}$ is required to achieve $\gamma \sim 2$ for the combination of baryonic and dark matter. However, cosmological hydrodynamic simulations have been unable to match both the observed $f_{\rm dm}$ and slope distribution. The $\gamma$ distribution can be reproduced in simulations by having no or weak feedback, but this leads to an overestimated galaxy formation efficiency and underestimated $f_{\rm dm}$ \citep{Naab07, Duffy10, Johansson12}. Conversely, reproducing the observed $f_{\rm dm}$ requires strong feedback, but then the predicted $\gamma$ distribution is too shallow. \citet{Dubois13} similarly find that simulation with AGN feedback can predict the observed $f_{\rm dm}$ in 0.4--8 $\times 10^{13}$ M$_{\odot}$ halos at $z=0$, but underestimate the $\gamma$ distribution. Without the AGN feedback, overestimated galaxy formation efficiency leads to underestimated $f_{\rm dm}$ and overestimated $\gamma$ distribution. More recently, \citet{Xu17} studied simulated elliptical galaxies from the Illustris hydrodynamic simulation that incorporates a number of baryonic processes \citep{Vogelsberger14}. These authors find higher $f_{\rm dm}$ and lower average $\gamma$ in Illustris galaxies than those observed in lens elliptical galaxies \citep{Auger09, Oldham18}. 
These mismatches 
point to either inadequacy in the theoretical model or systematic biases in the observational methods.

Many of the observational constraints on $f_{\rm dm}$ and $\gamma$ come from strong lensing systems with elliptical galaxies as deflector. Strong gravitational lensing provides a robust probe \editfv{of} the total projected mass within the Einstein radius. \editfv{Thus, combining} the stellar kinematics \editfv{with the lensing information can constrain the mass distribution in the deflector galaxy} \citep[e.g.,][]{Auger09, Sonnenfeld15}. To constrain $f_{\rm dm}$, decoupling the baryonic and dark components in the total mass is necessary. In previous studies, the stellar mass was inferred from the spectral energy distribution to decouple the dark and baryonic components \editth{\citep[e.g.,][]{Auger09, Spiniello11}}. This stellar mass depends on the assumption of the stellar initial mass function (IMF) introducing an uncertainty by a factor of $\sim$3. \editf{\editfv{However}, the IMF can be constrained \editfr{by making assumption on the mass profile \citep{Treu10}, or} by decoupling the baryonic and dark components with other external constraints \citep[e.g.,][]{Spiniello12, Barnabe13, Sonnenfeld19}.}

\editf{In this paper, we aim to constrain the $\gamma$ distribution and $f_{\rm dm}$ of elliptical galaxies \editfv{from the lensing and kinematics data -- independent of the SED-based stellar mass measurements --} and to constrain the \editfv{amount} of \editfv{adiabatic} contraction (or expansion) in these galaxies.} We model a sample of \numlens\ galaxy--galaxy lenses to study their structural properties. These lenses are assembled from the SLACS survey \citep{Bolton06, Bolton08}. 
\editfv{Previous SLACS analyses measured only the Einstein radius from the imaging data and then constrained the radially averaged logarithmic slope using stellar kinematics in combination with the imaging data.}
Due to several improvements in lens modelling techniques in the \editfv{past} decade, we can now constrain the logarithmic slope only from \editfv{the imaging} data, exploiting the richness of pixel-level information in the lensed arcs \citep[e.g.,][]{Suyu10b, Birrer15}. We model the SLACS lenses in our sample using state-of-the-art lens modelling techniques that simultaneously reconstruct the sources to extract the information contained in the lensed arcs. 
\editfv{Thus, we measure the local logarithmic slope at the Einstein radius only from the imaging data, independent of the stellar kinematics. We then combine the stellar kinematics with the lensing constraints to individually constrain the stellar and dark matter distributions, and infer the amount of adiabatic contraction in the dark matter distribution. Lensing-only measurement of the mass distribution is prone to the mass-sheet degeneracy \cite[MSD;][]{Falco85}. We adopt MSD-invariant quantities as the lensing constraints in our joint lensing--dynamics analysis, and combining these constraints with the stellar kinematics allows us to constrain the MSD in the inferred mass distribution.}


This paper is organized as follows. In Section \ref{sec:lens_sample}, we describe our lens sample and the imaging data. Next in Section \ref{sec:modelling}, we describe our uniform modelling procedure for this sample. We report the structural properties of elliptical lens galaxies from the lens models in Section \ref{sec:result}. We discuss our results in Section \ref{sec:discussion} and summarize the paper in Section \ref{sec:summary}. \editfr{Additionally in Appendix \ref{app:alignments}, we investigate the alignment between mass and light distributions.} We adopt a flat $\Lambda$ cold dark matter model as the fiducial cosmology with $H_0$=70 km s\textsuperscript{-1} Mpc\textsuperscript{-1}, $\Omega_{\rm m}=0.3$\editref{, and $\Omega_{\rm b} = 0.047$}. 
The reported uncertainties are obtained from 16\textsuperscript{th} and 84\textsuperscript{th} percentiles of the corresponding posterior probability distributions. \editet{We use $\log x$ to express the natural logarithm and $\log_{10}x$ to express the common logarithm.}

\section{Lens sample} \label{sec:lens_sample}

Our lens sample consists of \numlens\ galaxy--galaxy lenses from the SLACS survey. We first selected 50 galaxies from the full SLACS sample of 85 lenses by visually inspecting the lens images \editth{and selecting those}: (i) without nearby satellite or line-of-sight galaxies, (ii) without highly complex source morphology, (iii) with \textit{HST} imaging data in the \editfv{F555W/F606W bands (hereafter $V$ band)}, and (iv) not disc-like. Criteria (i) and (ii) are adopted so that we can uniformly apply our modelling procedure to the whole sample without needing to tweak the lens model settings on lens-by-lens basis. \editth{We adopt criterion (iii), because in the $V$ band the deflector galaxy is relatively fainter in comparison with the lensed arcs than in \editfv{the F814W band (hereafter $I$ band)}, which makes it easier to decouple the deflector light from the lensed arcs during lens modelling. We provide the list of selected galaxies in Appendix \ref{app:selected_galaxies}.}

\subsection{Imaging data}
Among the selected galaxies, some have imaging data from Advanced Camera for Surveys (ACS), and the rest from Wide Field and Planetary Camera 2 (WFPC2). The ACS images were taken with the F555W filter and the WFPC2 images were taken with the F606W filter. The images are obtained under the \textit{HST} GO programs 10494 (PI: Koopmans), 10798 (PI: Bolton), 10886 (PI: Bolton), and 11202 (PI: Koopmans).

The WFPC2 images were reduced for the original SLACS analysis \citep{Auger09}. We reduce the ACS images using the standard \textsc{astrodrizzle} software package \citep{Avila15}. The final pixel scale after drizzling is 0.05 arcsec.

We obtain the point spread function (PSF) for each filter and camera combination using \textsc{tinytim} \citep{Krist11}.

\subsection{Stellar kinematics data}
\editth{We use the line-of-sight velocity dispersions of the lenses in our sample measured from \editfv{the Sloan Digital Sky Survey (SDSS)} spectra. The fibre radius is 1.5 arcsec and the typical seeing for the observations is 1.4 arcsec. \citet{Bolton08} first measured the velocity dispersions from the SDSS reduction pipeline. \citet{Shu15} improved the measurements by updating the set of templates used to fit the spectra. We use this improved kinematics measurements in this study. These measurements are in good agreement with the Very Large Telescope (VLT) X-Shooter measurements of a subsample presented by \citet{Spiniello15}.}

\editth{\citet{Birrer20} find a residual scatter in the joint \editfv{lensing--dynamics} analysis using the same kinematics data used in this study, which accounts for $\sim$6 per cent unaccounted systematic uncertainty in the measured kinematics. Therefore, we add 6 per cent uncertainty in quadrature to the measured uncertainties.}

\subsection{Weak lensing data}

\editth{We incorporate weak lensing shear measurements of a sample of 33 SLACS lenses. The measurement pipeline is described by \citet{Gavazzi07}, and the sample size of the analysed SLACS lenses is increased by \citet{Auger10}. \citet{Sonnenfeld18b} find that the shear measurements of these 33 lenses is consistent with the shear measurement of the Hyper-Suprime Cam (HSC) survey weak lensing measurements after weighting the HSC sample to match the stellar mass distribution of the SLACS lenses.}

\editth{Out of the 23 SLACS lenses in our sample, 11 have directly measured reduced shear $\tilde{\gamma}_{\rm shear} \equiv \gamma_{\rm shear}/(1 - \kappa)$. For the remaining 12 lenses, we adopt the mean and scatter \editfv{(which includes both the intrinsic scatter and the noise)} of the measured reduced shears for the 33 lenses as the measured value and uncertainty, respectively. We adopt binned reduced shears only up to $\sim$100 kpc as the weak lensing constraint in our analysis. We do not use measurements beyond 100 kpc to avoid any potential bias from the 2-halo term as this term is not accounted for in our model. The centres of the adopted four bins are logarithmically spaced at 9.87 kpc,  17.78 kpc,  32.04 kpc, and  57.72 kpc.}

\section{Lens modelling} \label{sec:modelling}

We model the lenses using the lens modelling software \textsc{lenstronomy}, which is publicly available on GitHub\footnote{\faGithub\ \url{https://github.com/sibirrer/lenstronomy}} \citep{Birrer15, Birrer18}. \editref{The robustness of \textsc{lenstronomy} in recovering lens model parameters has been verified through Time-Delay Lens Modelling Challenge \citep[TDLMC;][]{Ding17, Ding20}. Two separate participating teams used \textsc{lenstronomy} to successfully recover the hidden Hubble constant and lens model parameters with statistical consistency for TDLMC Rung 2.} We package our modelling code into the \textsc{dolphin} pipeline\footnote{\faGithub\ \url{https://github.com/ajshajib/dolphin}}, which is a wrapper for \textsc{lenstronomy} to uniformly model large lens samples. First in Section \ref{sec:model_components}, we describe the components in our uniform lens model. Then in  Section \ref{sec:optimization}, we describe the optimization procedure for the lens model and Bayesian inference of the model parameters. \editfv{Next in Section \ref{sec:psf_effect}, we assess the effect of the PSF on the measured logarithmic slopes.}

\subsection{Model components} \label{sec:model_components}

We adopt the power-law ellipsoidal mass distribution (PEMD) for the deflector \citep{Barkana98}. \editsx{Although we aim to individually constrain the dark matter and stellar distributions from a joint lensing--dynamics analysis, first we only need to constrain MSD-invariant local lensing properties from our lens models, for which the PEMD model is sufficient. We combine these local lensing constraints with the stellar kinematics data in our joint lensing--dynamics analysis to individually constrain the dark matter and stellar distributions in Section \ref{sec:contraction}.} The convergence for the PEMD is given by
\begin{equation}
	\kappa (x, y) = \frac{3 - \gamma}{2}	\left( \frac{R_{\rm E}}{\sqrt{q_{\rm m} x^2 + y^2 / q_{\rm m}}} \right)^{\gamma-1},
\end{equation}
where $R_{\rm E}$ is the Einstein radius, $q$ is the axis ratio, and $\gamma$ is the logarithmic slope for the mass distribution in 3D. For an isothermal profile, the logarithmic slope is $\gamma=2$. The on-sky coordinates $(x,\ y)$ are rotated by position angle PA\textsubscript{m} from the (RA, \editref{DEC}) coordinates to align the $x$--axis with the major axis of the projected mass distribution. We also adopt an external shear profile parametrized with the shear magnitude $\gamma_{\rm ext}$ and the shear angle $\phi_{\rm ext}$.

We adopt a double S\'ersic profile for the deflector's light distribution, as a single S\'ersic profile leaves significant residual at the galaxy's centre \citep{Claeskens06, Suyu13}. The S\'ersic profile is given by
\begin{equation}
	I (x, y) = I_{\rm e} \exp	\left[ -k \left\{ \left( \frac{\sqrt{x^2 + y^2/q_{\rm L}^2}}{R_{\rm eff}} \right)^{1/n_{\rm s}} - 1 \right\} \right],
\end{equation}	
where $R_{\rm eff}$ is the effective radius, $I_{\rm e}$ is the surface brightness at $R_{\rm eff}$, $q_{\rm L}$ is axis ratio, $n_{\rm s}$ is the S\'ersic index, and $k$ is a normalizing constant so that $R_{\rm eff}$ becomes the half-light radius \citep{Sersic68}. The position angle for the light distribution is PA\textsubscript{L}. To constrain the degeneracy between the pairs of $R_{\rm eff}$ and $n_{\rm s}$ in the double S\'ersic profile, we fix $n_{\rm s} = 1$ and $n_{\rm s} = 4$, i.e., the exponential profile and the de Vaucouleurs profile respectively \citep{deVaucouleurs48}. For simplicity, we also join the ellipticity parameters \editth{$q_{\rm L}$ and PA$_{\rm L}$} between the two S\'ersic profiles.

We reconstruct the source galaxy's light distribution with a basis of shapelets and a S\'ersic profile \citep{Refregier03, Birrer15}. \editref{Similar shapelet-based source reconstructions have been successfully used to model a sample of 13 quadruply lensed quasars \citep{Shajib19}, and to model time-delay lenses to measure the Hubble constant \citep{Birrer19, Shajib20}.} The order parameter $n_{\rm max}$ determines the number of shapelets as $N_{\rm shapelets} = (n_{\rm max}+1)(n_{\rm max}+2)/2$. The \editf{scale size} of the shapelets is ruled by the scaling parameter $\beta$.

\subsection{Optimization and inference} \label{sec:optimization}

We obtain the posterior probability distributions for our model parameters using the Markov chain Monte Carlo (MCMC) method. If the MCMC sampling is started from a point close to the maxima of the posterior, the chain can converge with relatively less computational time. Therefore, we first optimize the lens model to get a point close to the maxima of the posterior. We use the particle swarm optimization (PSO) method for this step \citep{Kennedy95}. To further make this optimization computationally efficient, we adopt the following optimization recipe:

\begin{enumerate}[1.]
	\item Join the deflector mass and light centroids and fix the logarithmic slope $\gamma=2$ and shear magnitude $\gamma_{\rm ext}=0$ for all the steps below.
	\item Create a mask for the lensed arcs. We provide the algorithm to automatically make the mask for the arcs in Appendix \ref{app:arc_mask}. Fix all the model parameters except for the deflector's light profile. Optimize the deflector light parameters masking the lensed arcs.
	\item \editf{To find a good starting point for the source light parameters,} \editth{fix the lens model parameters and the deflector light parameters}. Fix the Einstein radius $R_{\rm E}$ and ellipticity parameters $\{q_{\rm m},\ \textrm{PA}_{\rm m}\}$ to the values measured by the SLACS analysis \citep{Auger09}. If such pre-determined values are not available, $R_{\rm E}$ can be fixed to an approximate guess and the ellipticity parameters can be set to the values for the circular case. Fix the shapelet scale parameter $\beta=$ 0.1 arcsec. Optimize only the remaining source light parameters to find an approximate position of the source on the source plane. Note, in this step the lensed arcs are not masked.
	\item Keep the deflector light parameters fixed and optimize for the source parameters and the PEMD parameters together. Keep $\beta=$ 0.1 arcsec fixed in this step.
	\item \editfv{Free} $\beta$ and optimize all the non-fixed parameters together.
	\item Repeat steps 2--5 \editth{with the current initial conditions from the previous step}.
\end{enumerate}

\editth{We notice prominent \editfv{residuals} at the centre of the deflector after performing the above automated procedure. \editfv{This is expected because the centres of elliptical galaxies are not perfectly described by S\'ersic profiles and our signal-to-noise ratio is very high in the centre.} To avoid bias in the model from this poor fitting of the deflector light profile at the centre, we mask out the central 0.4 arcsec and rerun the whole fitting procedure. The deflector light profile parameters for these systems are constrained from the light distribution that falls outside the central masked region. However for four systems -- J0252$+$0039, J1112$+$0826, J1313$+$4615, and J1636$+$4707 -- masking the deflector centre leads to even poorer quality fits \editfr{as evaluated with the p-value of the $\chi^2$ statistic}. Therefore, we do not mask the deflector centres for these four systems.}

\editth{We set $n_{\rm max} = 6$ for most of the lens systems in our sample. However, for the following systems we have adopted the following $n_{\rm max}$ values through trial-and-error with the above optimization procedure --  J0252$+$0039: 10, J0959$+$0410: 15, J1250$+$0523: 12, J1313$+$4615: 10, J1630$+$4520: 15.}

After the pre-sampling optimization, we then initiate the MCMC sampling from the  optimized lens model \editth{after step 6}. During the sampling, we \editfv{free} the parameters $\gamma$ and $\gamma_{\rm ext}$ that were fixed in the optimization step. We also independently sample the centroids of the deflector mass and light distributions. We perform the MCMC sampling using \textsc{emcee}, which is \editfv{an} affine-invariant ensemble sampler \citep{Goodman10, Foreman-Mackey13}. We assure the convergence of the chain by checking that the median and standard deviation of the \textsc{emcee} walkers at each step have \editth{reached equilibrium}.

\begin{figure*}
	\includegraphics[width=0.8\textwidth]{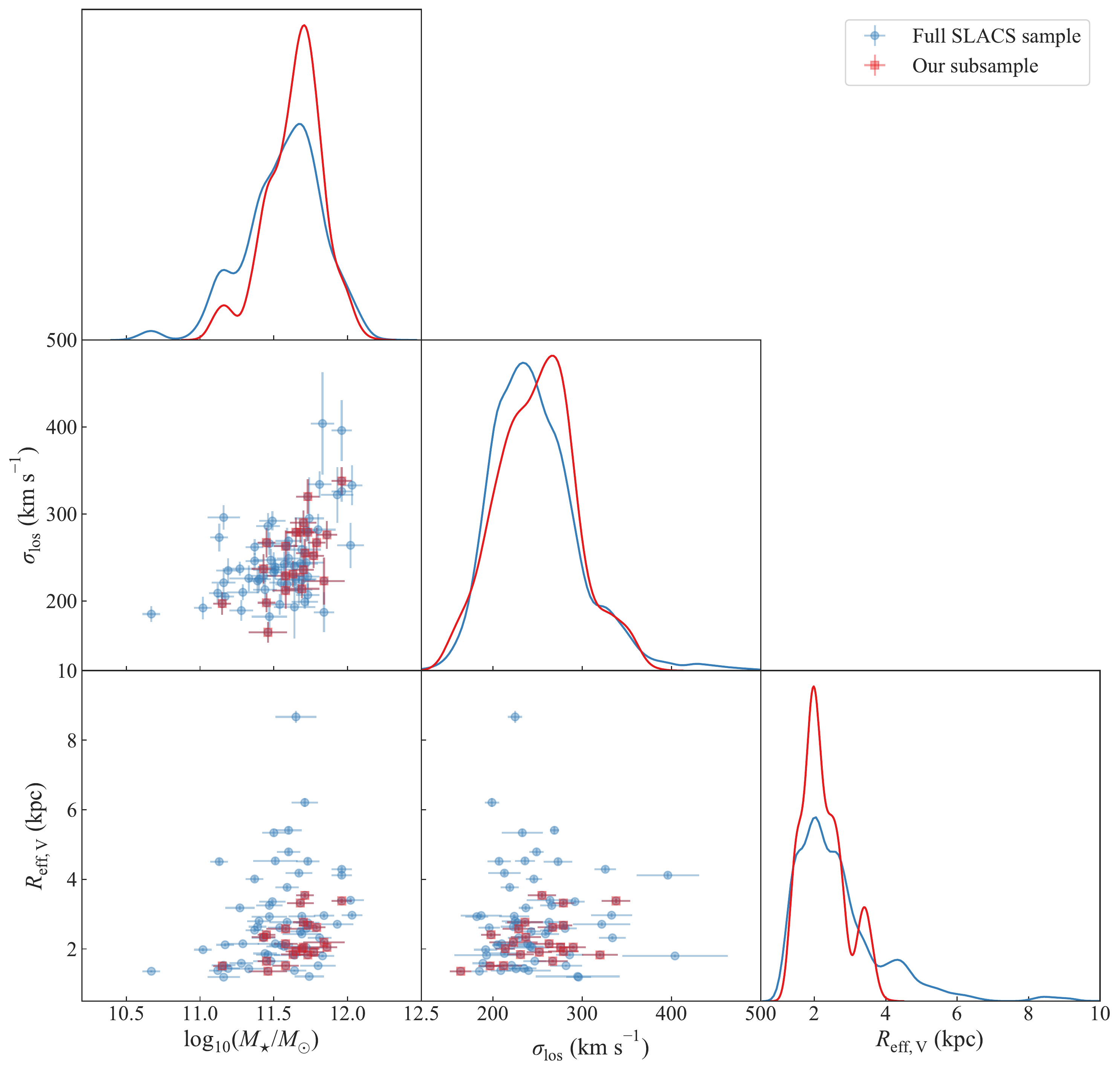}
	\caption{\label{fig:compare_sample}
	Comparison of the stellar mass, velocity dispersion, and the effective radius distributions between the full SLACS sample \citep[blue;][]{Auger09} and our selected subsample (red). 
		\editreft{The $p$-values from the 1D Kolmogorov--Smirnov test for each of the $\log_{10}({M}_{\star}/M_{\odot})$, $\sigma_{\rm los}$, and $R_{\rm eff, V}$ distributions are 0.37, 0.84, and 0.49, respectively. Therefore, the null hypothesis that our subsample is representative of the full subsample cannot be rejected with high statistical significance.} \editrefth{For reference, a $p$-value of 0.05 would allow us to reject the null hypothesis with 95 per cent confidence level.}
	}
\end{figure*}

\begin{figure*}
	\includegraphics[width=1.\textwidth]{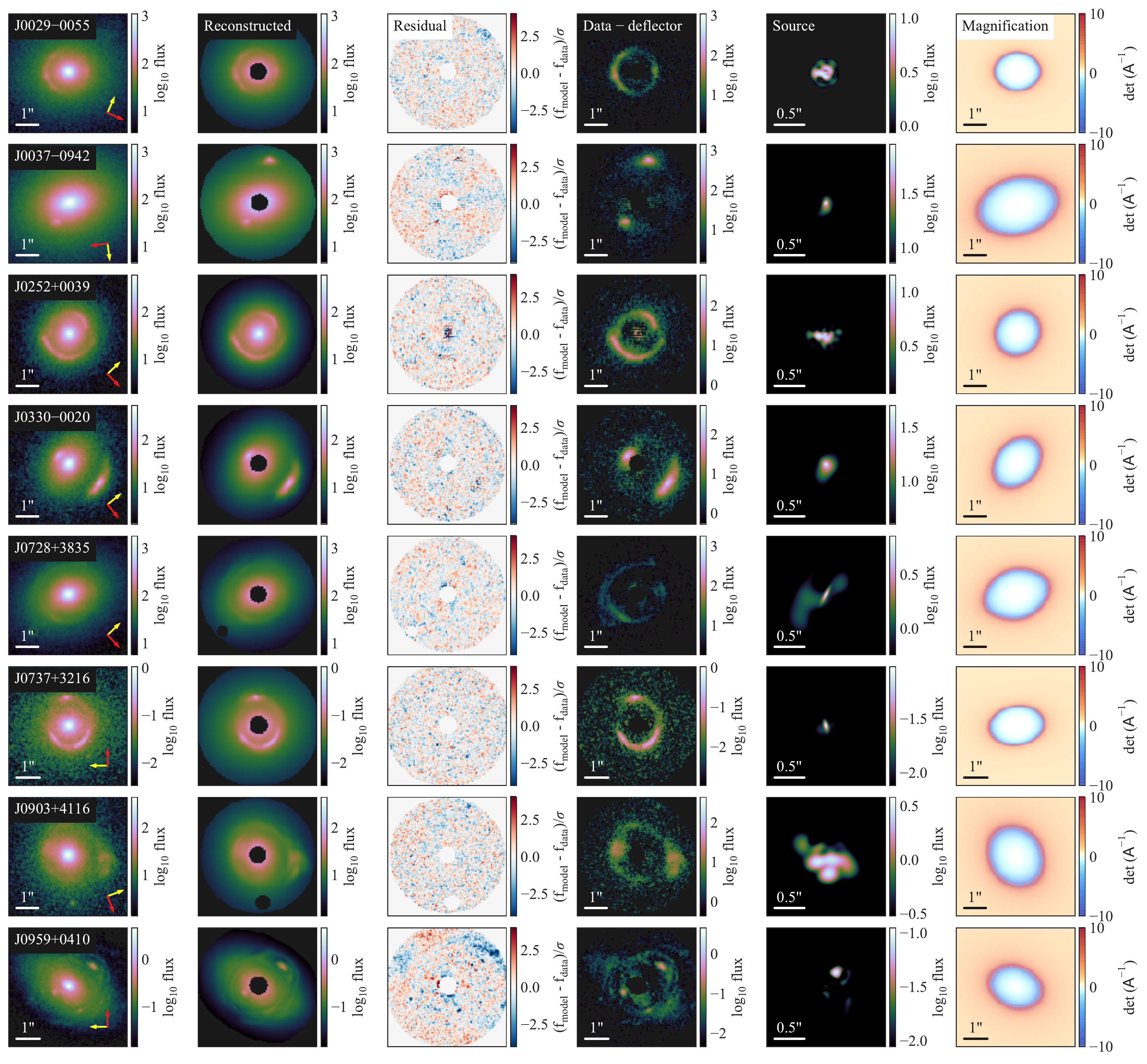}
	\caption{ \label{fig:montage_1}
	Lens models for \editf{the first 8 out of the \numlens} SLACS lenses in our sample. The first column shows the \textit{HST} images in F555W or F606W band. The second column shows the reconstructed image from the best-fitting lens model and the third column shows the normalized residual for the reconstruction. In the fourth column, we subtract the modelled deflector light distribution from the observed image to visualize the lensed arcs. The fifth column shows the reconstructed source and the sixth column shows the magnification map of the lens model. The red arrow points to the North and the yellow arrow points to the East. \editth{We mask the central 0.4 arcsec of some deflectors where there would otherwise be large residuals in the deflector's light profile fit without the mask.} \editf{The} lens models for the remaining systems are shown in Figures \ref{fig:montage_2} and \ref{fig:montge_3}.
	}
\end{figure*}

\begin{figure*}
	\includegraphics[width=1.\textwidth]{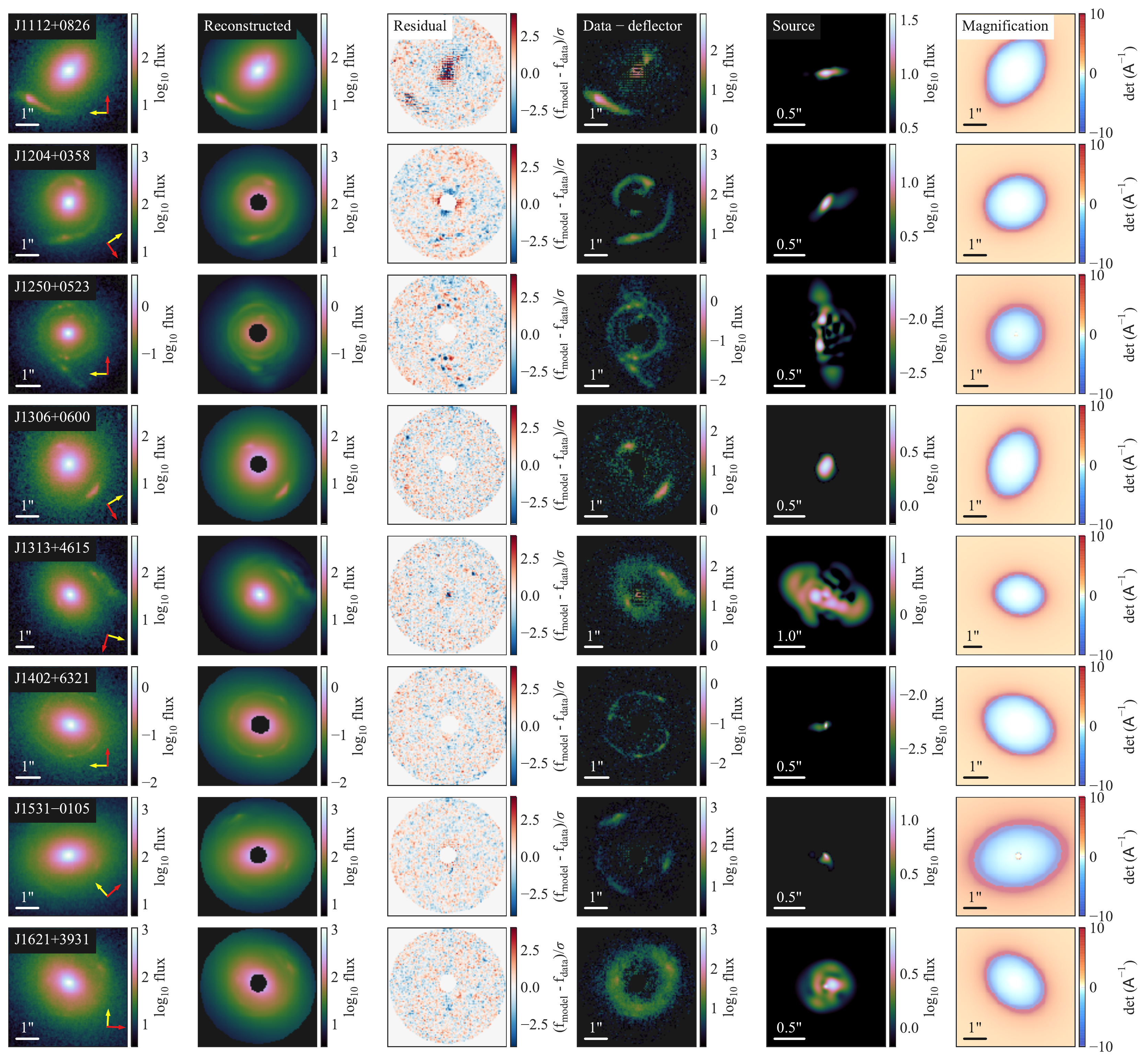}
	\caption{ \label{fig:montage_2}
	Lens models for the next 8 out of the \numlens\ SLACS lenses continuing after Figure \ref{fig:montage_1}. 
	}
\end{figure*}

\begin{figure*}
	\includegraphics[width=1.\textwidth]{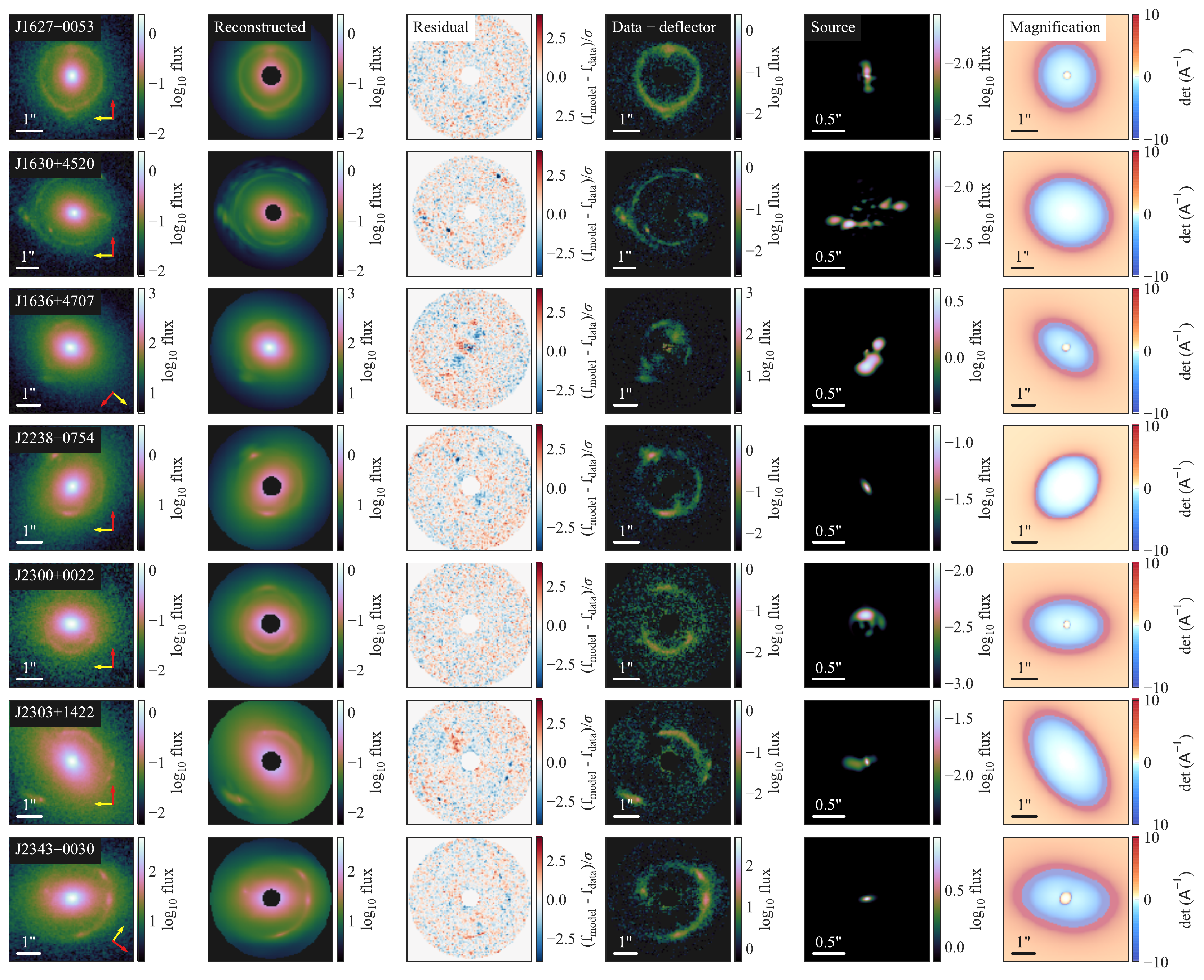}
	\caption{ \label{fig:montge_3}
	Lens models for the last 7 out of the \numlens\ SLACS lenses continuing after Figures  \ref{fig:montage_1} and \ref{fig:montage_2}. 
	}
\end{figure*}

After our uniform modelling procedure for the 50 \editth{initially} selected lenses, we vet for reliability of the lens models based on the \editth{following} criteria: (i) absence of prominent model residuals indicative of poor source reconstruction, and (ii) \editfr{the median of the inferred logarithmic slope has not converged to too low ($\gamma\lesssim1.4$) or too high ($\gamma \gtrsim 2.8)$ values. The $\gamma$ distribution can converge towards such extreme values if the lensed arcs are faint and thus the lensing information contained in the imaging data is not sufficient to constrain $\gamma$. When $\gamma$ is too small ($\gamma \lesssim 1.4$), a central image will be produced in the lens models, which is not observed in the imaging data by looking at the color distribution. However, since we mask out the central region for most of our lens systems, a central image is not directly penalized in the likelihood term. Moreover, since the lensed arcs are relatively fainter, prominent residuals are not noticeable in the lens models even if the source reconstruction is poor. We treat such extreme values of $\gamma$ as numerical artefacts due to weak constraining power of the imaging data and remove these systems from our sample. Note that criterion (ii) is practically a uniform prior ${\rm median} (\gamma) \sim \mathcal{U}(1.4,\ 2.8)$.} 
After this vetting procedure, we are left with \numlens\ systems with reliable lens models. \editref{This subsample of 23 lenses is representative of the full sample of 85 SLACS lenses in terms of the stellar mass and velocity dispersion distributions, however our selection excludes the lens galaxies with relatively larger effective radii ($R_{\rm eff, V} \gtrsim 5$ kpc; Figure \ref{fig:compare_sample}).} \editreft{We perform the 1D Kolmogorov--Smirnov test for each of the $\log_{10}({M}_{\star}/M_{\odot})$, $\sigma_{\rm los}$, and $R_{\rm eff, V}$ distributions between the full SLACS sample and our subsample. The $p$-values are 0.37, 0.84, and 0.49, respectively. Therefore, the null hypothesis that our subsample is representative of the full SLACS sample cannot be rejected with high significance.} \editrefth{For reference, a $p$-value of 0.05 would allow us to rule out the null hypothesis with 95 per cent confidence level.} We show the lens images and the models in Figures \ref{fig:montage_1}, \ref{fig:montage_2}, and \ref{fig:montge_3}. \editfv{We tabulate the marginalized posteriors of the lens model parameters in Table \ref{tab:lens_params}.}

\subsection{Effect of PSF on measured logarithmic slope} \label{sec:psf_effect}

\begin{figure*}
	\includegraphics[width=1.\textwidth]{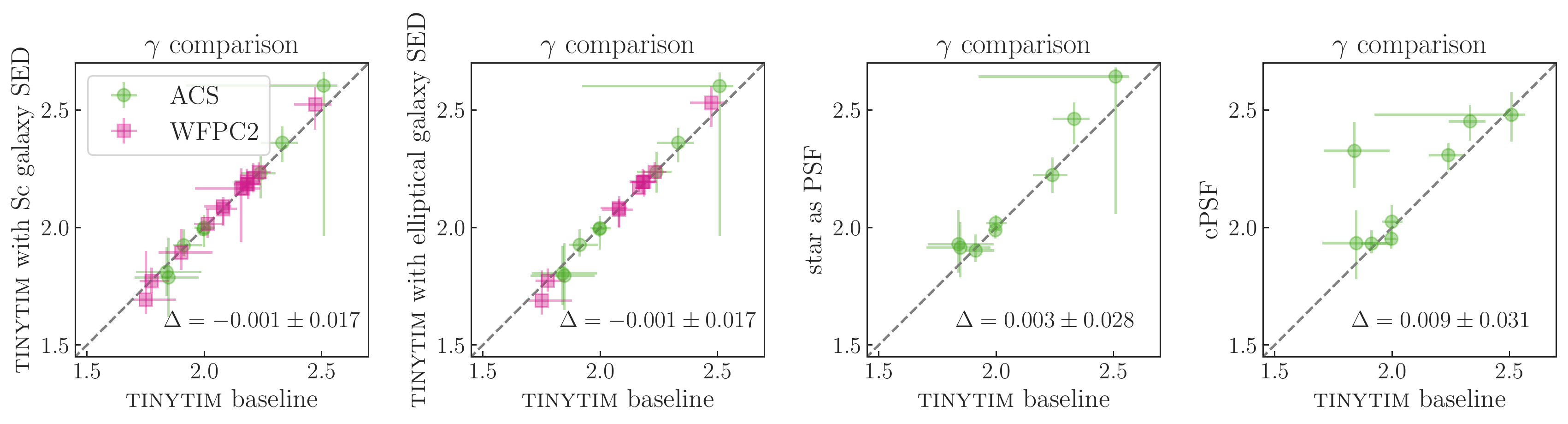}
	\caption{ \label{fig:psf_compare}
	Comparison of measured logarithmic slopes for various choices of PSF. The \editref{five} choices are: (i) \textsc{tinytim} PSF with G2V star SED (baseline), (ii) \textsc{tinytim} PSF with Sc galaxy SED at $\langle z_{\rm s}\rangle=0.62$, (iii) \textsc{tinytim} PSF with elliptical galaxy SED at $\langle z_{\rm d} \rangle = 0.2$,
	(iv) centreed and stacked stars from the corresponding \textit{HST} image, and (v) ePSF created from the stars in the corresponding \textit{HST} image \citep{Anderson00}. The systems with ACS F555W imaging are shown with green circles and the systems with WFPC2 F606W imaging are shown with pink squares. Reliable PSFs for PSF types (iv) and (v) could be extracted only for the ACS images in our sample. The \editf{mean difference} $\Delta$ between two corresponding PSF types is annotated within each panel. In all cases, the deviation of the sample mean is negligible relative to the uncertainty of individual logarithmic slopes (typically $\gtrsim 0.04$).
	}
\end{figure*}

We check the effect of the PSF choice on the measured logarithmic slopes. We adopt the following five PSF choices for this test:
\begin{enumerate}[(i)]
	\item \textsc{tinytim} PSF with G2V star SED as our baseline PSF,
	\item \editth{\textsc{tinytim} PSF with Sc galaxy SED at redshift $\langle z_{\rm s} \rangle =0.62$, which is the average source redshift in our sample,}
	\item \editth{\textsc{tinytim} PSF with elliptical galaxy SED at redshift $\langle z_{\rm d} \rangle=0.20$, which is the average deflector redshift in our sample,}
	\item a PSF created from re-centreing and stacking star cut-outs from the corresponding \textit{HST} image, and
	\item effective PSF (ePSF) created from the stars from the corresponding \textit{HST} image \citep{Anderson00}.
\end{enumerate}
%
For PSF choices (iv) and (v), reliable PSFs could be extracted only for the ACS images. \editf{Therefore, we only check with PSF choice (iv) and (v) for nine systems with ACS images from our sample. We re-optimize our lens models for PSF types (ii) to (v).} We compare the measured logarithmic slopes with these PSF choices in Figure \ref{fig:psf_compare}. The mean deviation $|\Delta| \sim 0.001$--$0.009$ of the logarithmic slope distribution is negligible relative to the uncertainty of the individual logarithmic slope (typically $\sigma_{\rm \gamma} \gtrsim 0.04$) when the SED of the PSF is varied. 
\editfv{However}, a scatter of approximately 0.02--0.03 is introduced in the logarithmic slope distribution from the choice of PSF.

\section{Structural properties of lens galaxies} \label{sec:result}
In this section, we report our findings on the structural properties of the lens galaxies. 
In Section \ref{sec:slope_dist}, \editsx{we present the distribution of the logarithmic slopes constrained from the lensing-only data}.  
\editfv{Next} in Section \ref{sec:contraction}, we combine stellar kinematics with the lensing observables to infer the \editfv{amount} of contraction in our sample.

\subsection{The distribution of logarithmic slopes} \label{sec:slope_dist}

\begin{figure}
	\includegraphics[width=\columnwidth]{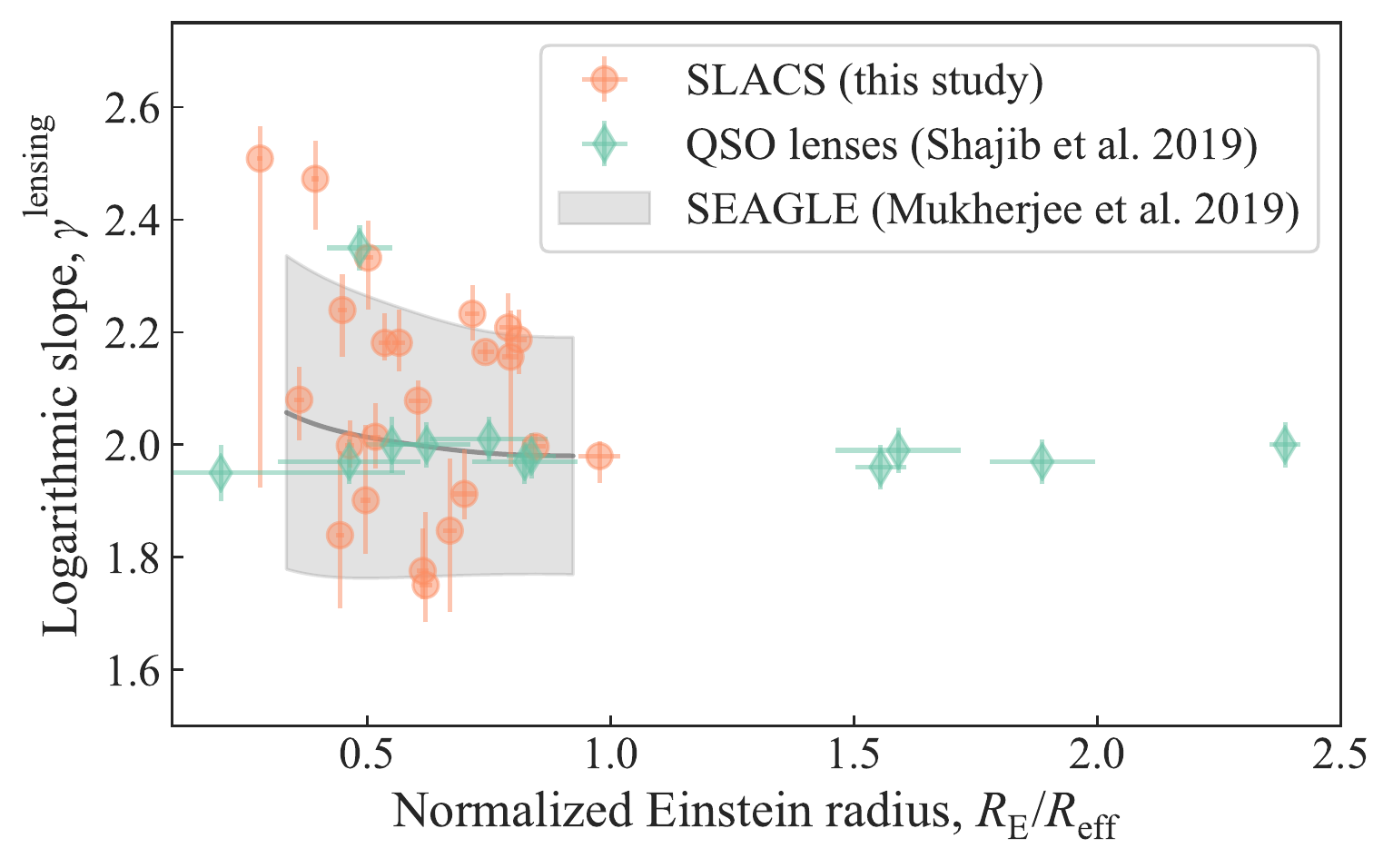}
	\caption{ \label{fig:slope_dist}
	Distribution of the local logarithmic slopes for the SLACS lenses (orange points). We take $R_{\rm E}/R_{\rm eff}$ along the $x$-axis to normalize the \editfr{radius where the local slope is estimated.} 
	For comparison, we show the distribution of slopes for strongly lensed quasars systems (green) from \citet{Shajib19}. \editfr{The grey solid line shows the mean of the slope distribution of simulated lens systems from SEAGLE and the grey shaded region shows 1 standard deviation around the mean \citep{Mukherjee19}. The \editfv{1$\sigma$ interval} is produced by smoothing the simulated points from SEAGLE AGNdT8 simulation -- which includes AGN feedback -- with a Gaussian kernel of bandwidth 0.15. Our measured values of the logarithmic slopes from lensing-only data match well with the distribution of similarly measured logarithmic slopes of simulated lens systems that include AGN feedback.}  
	}
\end{figure}

We plot the distribution of the SLACS lenses in our sample on the $\gamma$--$R_{\rm E}/R_{\rm eff}$ plane in Figure \ref{fig:slope_dist} left-hand panel. For comparison, we also plot the distribution of the 13 strongly lensed quasar systems from \citet{Shajib19}. Strong lensing constrains the slope of the projected mass profile at the Einstein radius. Therefore, if we assume that elliptical galaxies are self-similar, then this plot illustrates the distribution of slopes of the projected mass density at different scale sizes. The median of the distribution is $2.09^{+0.03}_{-0.04}$ and the intrinsic scatter of the distribution is $0.13 \pm 0.02$. We account for the uncertainty in individually measured $\gamma$ by sampling 1000 sets of $\gamma$'s assuming Gaussian uncertainty for each measurement and then obtaining the distribution of the medians and scatters from these 1000 sets.

\begin{figure}
	\includegraphics[width=\columnwidth]{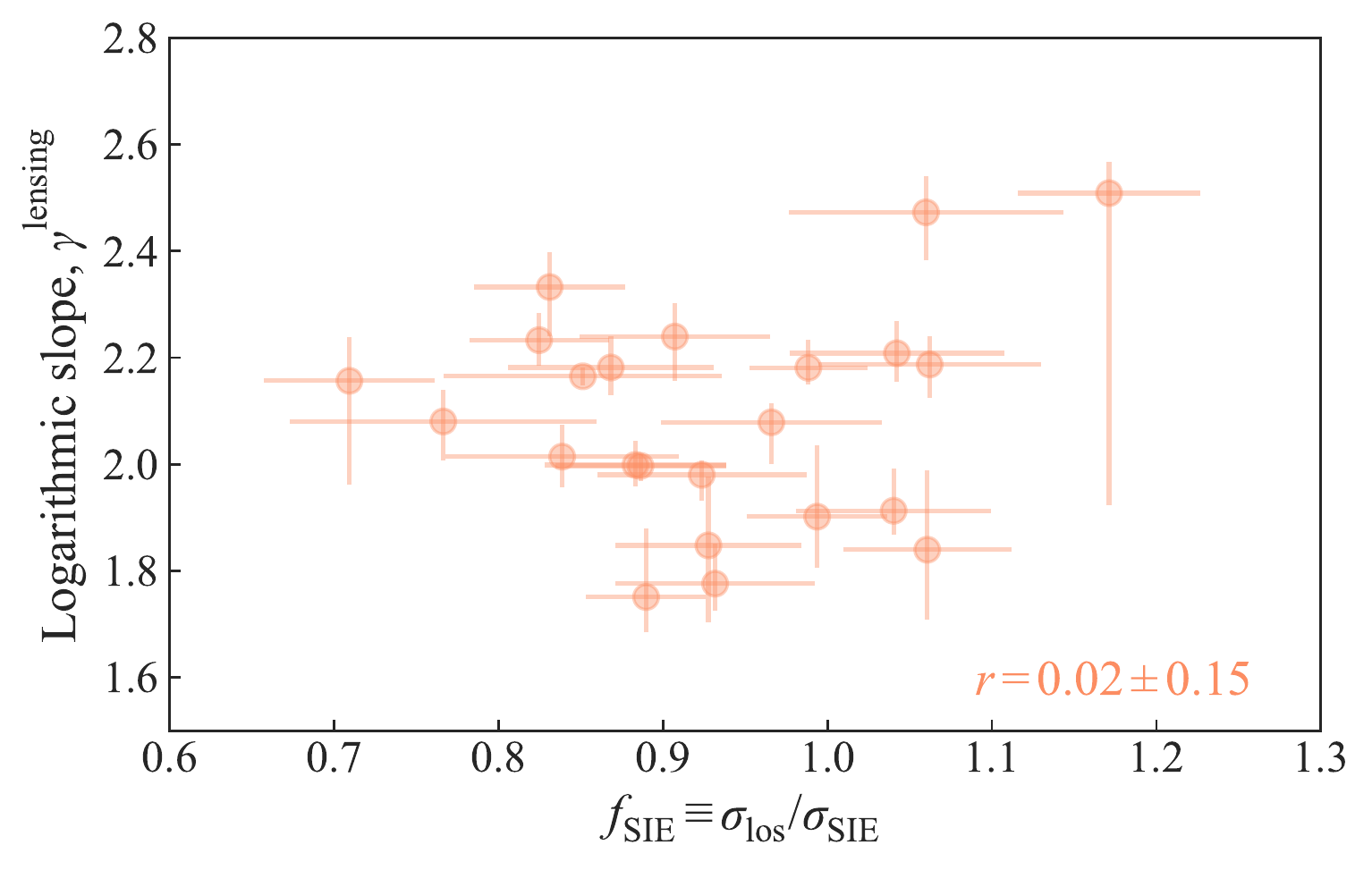}
	\caption{ \label{fig:slope_vs_f_sie}
	Distribution of $\gamma^{\rm lensing}$ and $f_{\rm SIE} \equiv \sigma_{\rm los}/\sigma_{\rm SIE}$ for the SLACS sample. The biweight mid-correlation $r$ between these two parameters is annotated  in the figure. The measured logarithmic slopes have no correlation with the $f_{\rm SIE}$. \editf{\editfr{If the global mass distribution follows the power law, then $\gamma$ should positively correlate with $f_{\rm SIE}$ [see equation (2.3) of \citet{Koopmans04}, cf. Figure 4 of \citet{Auger10b}]}. However, we find that the uncertainty level in the measured $\gamma$ and $f_{\rm SIE}$ are \ttedit{too large for us to detect a potential correlation. As a result, this comparison is inconclusive, and it cannot confirm or rule out whether the global mass distribution is well approximated by a power law. A more sophisticated analysis, such as that presented later in this paper, or better data are required for this investigation.}}
	}

\end{figure}

We find no correlation between observed $\gamma$ distribution and ratio $f_{\rm SIE} \equiv \sigma_{{\rm los}} / \sigma_{\rm SIE}$ between the observed velocity dispersion and the SIE velocity dispersion (Figure \ref{fig:slope_vs_f_sie}). \editth{We calculate the SIE velocity dispersion is given by
\begin{equation}
	\sigma_{\rm SIE} = c \sqrt{\frac{R_{\rm E} D_{\rm s}}{4 \uppi D_{\rm ds}}},
\end{equation}
where $c$ is the speed of light.} \editf{If the underlying true mass profile follows a power law, then $\gamma$ should positively correlate with $f_{\rm SIE}$ [see equation (2.3) of \citet{Koopmans04}, cf. Figure 4 of \citet{Auger10b}]. Therefore, there are two possible explanations for this lack of correlation: (i) the underlying true mass profile deviates from a power-law, and/or (ii) the noise is large enough that any potential correlation is washed away. We perform a linear regression for the $\gamma$--$f_{\rm SIE}$ distribution with intrinsic scatter as a free parameter. We find that only $3.5$ per cent or less intrinsic scatter is necessary with 95 per cent confidence to fit the data. This result indicates that the uncertainties in $\gamma$ and $f_{\rm SIE}$ are too large \ttedit{to detect the expected correlation for a power-law profile}.} \editfv{This is a case of absence of evidence, not evidence of absence.}

\renewcommand{\arraystretch}{1.2}
\begin{table*}
	\caption{\label{tab:lens_params}
		\editfr{Point estimates of the parameters in our power-law lens models. Here, $R_{\rm E}$ is the Einstein radius, $\gamma^{\rm lensing}$ is the logarithmic slope of the mass profile, $q_{\rm m}$ is the mass axis ratio, PA\textsubscript{m} is the mass position angle, $\gamma_{\rm ext}$ is the external shear magnitude, $\phi_{\rm ext}$ is the external shear angle, $R_{\rm eff}$ is the effective or half-light radius in $V$ band, $q_{\rm L}$ is the light axis ratio, and PA\textsubscript{L} is the light position angle.}
	}
	\begin{tabular}{lccccccccc}
	\hline
	Name & $R_{\rm E}$ & $\gamma^{\rm lensing}$ & $q_{\rm m}$ & PA\textsubscript{m} (N of E) &  $\gamma_{\rm ext}$ & $\phi_{\rm ext}$ (N of E) & \editfr{$R_{\rm eff}$} & $q_{\rm L}$ & PA\textsubscript{L} (N of E) \\
	& (arcsec) & & & (deg) & & (deg) & \editfr{(arcsec)} & & (deg) \\
	\hline 
J0029$-$0055 & $0.951_{-0.003 }^{ +0.003 }$ & $2.47_{-0.09 }^{ +0.07 }$ & $0.69_{-0.04 }^{ +0.04 }$ & $66_{-2 }^{ +2 }$ & $0.002_{-0.001 }^{ +0.002 }$ & $-22_{-63 }^{ +79 }$ & 2.42$\pm$0.05 & $0.837_{-0.002 }^{ +0.002 }$ & $62.5_{-0.5 }^{ +0.4 }$ \\
J0037$-$0942 & $1.509_{-0.029 }^{ +0.009 }$ & $2.18_{-0.03 }^{ +0.05 }$ & $0.65_{-0.02 }^{ +0.02 }$ & $82_{-3 }^{ +3 }$ & $0.013_{-0.009 }^{ +0.010 }$ & $21_{-13 }^{ +10 }$ & 2.80$\pm$0.06 & $0.701_{-0.001 }^{ +0.001 }$ & $82.9_{-0.1 }^{ +0.1 }$ \\
J0252$+$0039 & $1.011_{-0.002 }^{ +0.005 }$ & $2.16_{-0.20 }^{ +0.08 }$ & $0.89_{-0.03 }^{ +0.01 }$ & $-10_{-9 }^{ +8 }$ & $0.009_{-0.006 }^{ +0.017 }$ & $-33_{-23 }^{ +16 }$ & 1.27$\pm$0.03 & $0.931_{-0.005 }^{ +0.004 }$ & $-24.6_{-2.3 }^{ +2.1 }$ \\
J0330$-$0020 & $1.077_{-0.005 }^{ +0.005 }$ & $2.16_{-0.02 }^{ +0.02 }$ & $0.69_{-0.04 }^{ +0.04 }$ & $-7_{-3 }^{ +3 }$ & $0.025_{-0.008 }^{ +0.007 }$ & $26_{-11 }^{ +9 }$ & 1.45$\pm$0.03 & $0.738_{-0.004 }^{ +0.004 }$ & $-18.7_{-0.4 }^{ +0.4 }$ \\
J0728$+$3835 & $1.283_{-0.005 }^{ +0.004 }$ & $2.23_{-0.05 }^{ +0.05 }$ & $0.59_{-0.03 }^{ +0.03 }$ & $24_{-1 }^{ +1 }$ & $0.061_{-0.007 }^{ +0.007 }$ & $16_{-3 }^{ +3 }$ & 1.79$\pm$0.04 & $0.754_{-0.002 }^{ +0.002 }$ & $22.6_{-0.3 }^{ +0.3 }$ \\
J0737$+$3216 & $0.973_{-0.004 }^{ +0.015 }$ & $2.51_{-0.58 }^{ +0.06 }$ & $0.68_{-0.04 }^{ +0.04 }$ & $-5_{-4 }^{ +2 }$ & $0.117_{-0.106 }^{ +0.009 }$ & $78_{-31 }^{ +1 }$ & 3.49$\pm$0.07 & $0.858_{-0.006 }^{ +0.007 }$ & $-11.4_{-1.4 }^{ +1.2 }$ \\
J0903$+$4116 & $1.271_{-0.007 }^{ +0.014 }$ & $2.08_{-0.07 }^{ +0.06 }$ & $0.72_{-0.02 }^{ +0.12 }$ & $65_{-117 }^{ +3 }$ & $0.089_{-0.088 }^{ +0.009 }$ & $50_{-51 }^{ +3 }$ & 3.53$\pm$0.07 & $0.855_{-0.008 }^{ +0.006 }$ & $-86.8_{-1.2 }^{ +13.5 }$ \\
J0959$+$0410 & $0.962_{-0.017 }^{ +0.042 }$ & $1.98_{-0.05 }^{ +0.03 }$ & $0.71_{-0.02 }^{ +0.09 }$ & $13_{-2 }^{ +26 }$ & $0.058_{-0.057 }^{ +0.010 }$ & $-11_{-3 }^{ +11 }$ & 0.99$\pm$0.02 & $0.668_{-0.042 }^{ +0.033 }$ & $31.3_{-4.2 }^{ +4.5 }$ \\
J1112$+$0826 & $1.421_{-0.012 }^{ +0.017 }$ & $2.21_{-0.05 }^{ +0.06 }$ & $0.56_{-0.03 }^{ +0.06 }$ & $-57_{-1 }^{ +2 }$ & $0.066_{-0.037 }^{ +0.017 }$ & $-63_{-7 }^{ +173 }$ & 1.80$\pm$0.04 & $0.746_{-0.001 }^{ +0.001 }$ & $-46.5_{-0.2 }^{ +0.2 }$ \\
J1204$+$0358 & $1.285_{-0.004 }^{ +0.005 }$ & $2.19_{-0.06 }^{ +0.05 }$ & $0.82_{-0.03 }^{ +0.03 }$ & $-4_{-3 }^{ +3 }$ & $0.056_{-0.006 }^{ +0.005 }$ & $-41_{-4 }^{ +5 }$ & 1.59$\pm$0.03 & $0.987_{-0.003 }^{ +0.003 }$ & $-48.7_{-4.8 }^{ +5.2 }$ \\
J1250$+$0523 & $1.115_{-0.003 }^{ +0.005 }$ & $1.91_{-0.04 }^{ +0.08 }$ & $0.91_{-0.03 }^{ +0.02 }$ & $-60_{-4 }^{ +23 }$ & $0.020_{-0.019 }^{ +0.007 }$ & $-73_{-6 }^{ +74 }$ & 1.60$\pm$0.03 & $0.893_{-0.016 }^{ +0.006 }$ & $-5.2_{-1.4 }^{ +1.3 }$ \\
J1306$+$0600 & $1.299_{-0.013 }^{ +0.011 }$ & $2.18_{-0.05 }^{ +0.06 }$ & $0.66_{-0.07 }^{ +0.10 }$ & $-39_{-3 }^{ +3 }$ & $0.006_{-0.004 }^{ +0.008 }$ & $11_{-105 }^{ +58 }$ & 2.30$\pm$0.05 & $0.909_{-0.003 }^{ +0.003 }$ & $-50.2_{-0.9 }^{ +0.8 }$ \\
J1313$+$4615 & $1.302_{-0.014 }^{ +0.011 }$ & $2.08_{-0.08 }^{ +0.04 }$ & $0.64_{-0.03 }^{ +0.19 }$ & $-13_{-3 }^{ +35 }$ & $0.098_{-0.097 }^{ +0.011 }$ & $-20_{-3 }^{ +20 }$ & 2.15$\pm$0.04 & $0.806_{-0.048 }^{ +0.004 }$ & $30.2_{-0.6 }^{ +2.2 }$ \\
J1402$+$6321 & $1.354_{-0.003 }^{ +0.003 }$ & $2.24_{-0.08 }^{ +0.06 }$ & $0.70_{-0.04 }^{ +0.04 }$ & $24_{-2 }^{ +2 }$ & $0.010_{-0.005 }^{ +0.006 }$ & $1_{-19 }^{ +11 }$ & 3.02$\pm$0.06 & $0.758_{-0.002 }^{ +0.002 }$ & $17.9_{-0.3 }^{ +0.3 }$ \\
J1531$-$0105 & $1.705_{-0.011 }^{ +0.006 }$ & $1.90_{-0.10 }^{ +0.13 }$ & $0.68_{-0.03 }^{ +0.02 }$ & $-61_{-3 }^{ +4 }$ & $0.046_{-0.031 }^{ +0.012 }$ & $-69_{-8 }^{ +169 }$ & 3.43$\pm$0.07 & $0.672_{-0.001 }^{ +0.001 }$ & $-55.1_{-0.1 }^{ +0.1 }$ \\
J1621$+$3931 & $1.263_{-0.005 }^{ +0.004 }$ & $2.01_{-0.06 }^{ +0.06 }$ & $0.77_{-0.03 }^{ +0.02 }$ & $-50_{-3 }^{ +2 }$ & $0.007_{-0.004 }^{ +0.007 }$ & $41_{-150 }^{ +38 }$ & 2.44$\pm$0.05 & $0.739_{-0.002 }^{ +0.002 }$ & $-53.2_{-0.3 }^{ +0.3 }$ \\
J1627$-$0053 & $1.227_{-0.002 }^{ +0.002 }$ & $1.84_{-0.13 }^{ +0.15 }$ & $0.87_{-0.01 }^{ +0.01 }$ & $80_{-2 }^{ +2 }$ & $0.026_{-0.007 }^{ +0.007 }$ & $-96_{-5 }^{ +178 }$ & 2.76$\pm$0.06 & $0.836_{-0.004 }^{ +0.004 }$ & $85.9_{-0.6 }^{ +0.7 }$ \\
J1630$+$4520 & $1.788_{-0.006 }^{ +0.005 }$ & $2.00_{-0.03 }^{ +0.02 }$ & $0.83_{-0.01 }^{ +0.02 }$ & $20_{-4 }^{ +2 }$ & $0.019_{-0.018 }^{ +0.004 }$ & $26_{-26 }^{ +5 }$ & 2.11$\pm$0.04 & $0.825_{-0.044 }^{ +0.003 }$ & $19.7_{-0.9 }^{ +0.8 }$ \\
J1636$+$4707 & $1.100_{-0.007 }^{ +0.006 }$ & $1.78_{-0.05 }^{ +0.07 }$ & $0.66_{-0.04 }^{ +0.03 }$ & $-4_{-3 }^{ +2 }$ & $0.055_{-0.017 }^{ +0.014 }$ & $-3_{-7 }^{ +6 }$ & 1.79$\pm$0.04 & $0.797_{-0.001 }^{ +0.002 }$ & $-14.2_{-0.3 }^{ +0.3 }$ \\
J2238$-$0754 & $1.272_{-0.003 }^{ +0.002 }$ & $2.33_{-0.09 }^{ +0.07 }$ & $0.70_{-0.03 }^{ +0.03 }$ & $-44_{-1 }^{ +2 }$ & $0.003_{-0.002 }^{ +0.003 }$ & $-19_{-39 }^{ +39 }$ & 2.53$\pm$0.05 & $0.784_{-0.002 }^{ +0.002 }$ & $-50.7_{-0.3 }^{ +0.3 }$ \\
J2300$+$0022 & $1.243_{-0.006 }^{ +0.008 }$ & $1.85_{-0.14 }^{ +0.13 }$ & $0.69_{-0.04 }^{ +0.04 }$ & $5_{-2 }^{ +2 }$ & $0.025_{-0.018 }^{ +0.020 }$ & $17_{-12 }^{ +25 }$ & 1.86$\pm$0.04 & $0.790_{-0.003 }^{ +0.003 }$ & $4.0_{-0.5 }^{ +0.5 }$ \\
J2303$+$1422 & $1.611_{-0.005 }^{ +0.010 }$ & $2.00_{-0.04 }^{ +0.05 }$ & $0.59_{-0.03 }^{ +0.03 }$ & $54_{-1 }^{ +2 }$ & $0.002_{-0.002 }^{ +0.003 }$ & $1_{-17 }^{ +18 }$ & 3.48$\pm$0.07 & $0.659_{-0.002 }^{ +0.002 }$ & $52.5_{-0.2 }^{ +0.2 }$ \\
J2343$-$0030 & $1.529_{-0.005 }^{ +0.004 }$ & $1.75_{-0.07 }^{ +0.13 }$ & $0.65_{-0.03 }^{ +0.02 }$ & $54_{-1 }^{ +1 }$ & $0.088_{-0.007 }^{ +0.005 }$ & $34_{-8 }^{ +3 }$ & 2.46$\pm$0.05 & $0.662_{-0.005 }^{ +0.003 }$ & $51.1_{-0.2 }^{ +0.2 }$ \\
	\hline
	\end{tabular}

\end{table*}

\subsection{Dark matter contraction} \label{sec:contraction}

\editf{When gas inside a dark matter halo cools and condensates at the centre as stars begin to form, the dark matter distribution also contracts in response. If the gas infall is slow and smooth, then the dark matter contraction is adiabatic \citep{Blumenthal86}. Inversely, dark matter halo can also adiabatically expand, if there is gas outflow.}

We infer the contraction in the dark matter distribution in our lens galaxies by combining the lensing observables from our models with their measured stellar kinematics. \editth{For the dark matter distribution, we allow adiabatic contraction or expansion from the `pristine' NFW distribution.} We follow the contraction formalism of \citet{Dutton07}, which is originally based on the formalism of \citet{Blumenthal86}. In this formalism, the initial radius $r_{\rm i}$ of a dark matter particle and its final radius $r_{\rm f}$ after contraction is related as
\begin{equation} \label{eq:contraction_1}
	r_{\rm i} M_{\rm i} (r_{\rm i}) = r_{\rm f} M_{\rm f} (r_{\rm f}),
\end{equation}
where $M(r)$ is the total 3D mass within radius $r$. 
Then, we have 
\begin{equation} \label{eq:contraction_2}
	\begin{aligned}
		M_{\rm f} (r_{\rm f}) &= M_{\rm gal, f} (r_{\rm f}) + M_{\rm dm, f} (r_{\rm f}) \\
		&= M_{\rm gal, f} (r_{\rm f}) + (1 -f_{\rm gal}) M_{\rm i} (r_{\rm i}),
	\end{aligned}
\end{equation}
where $f_{\rm gal} \equiv M_{\rm gal,i}/M_{\rm i} = M_{\rm gal,i}/(M_{\rm dm,i} + M_{\rm gal,i})$ is the fraction of the total mass that cools to form stars. We iteratively solve equations (\ref{eq:contraction_1}) and (\ref{eq:contraction_2}) to obtain the contraction factor $\Gamma (r_{\rm f}) \equiv r_{\rm f} / r_{\rm i}$. Following \citet{Dutton07}, we furthermore adopt the halo response parameter $\nu$ to modify the relation between $r_{\rm i}$ and $r_{\rm f}$ as
\begin{equation}
	r_{\rm i} \equiv \Gamma^{-\nu} (r_{\rm f})\ r_{\rm f}.
\end{equation}
By varying $\nu$ in our models we can adjust the amount of contraction. For example, $\nu=0$ corresponds to no contraction, $\nu=1$ corresponds to full contraction as in \citet{Blumenthal86}, and $\nu<0$ corresponds to the case of adiabatic expansion. Cosmological hydrodynamical simulations have found less contraction than the model of \citet{Blumenthal86}, e.g., the result of \citet{Gnedin04} corresponds to $\nu = 0.8$ and the result of \citet{Abadi10} corresponds to $\nu \sim 0.4$ \citep{Dutton14}. \editsx{However, these simulations almost always find contracted dark matter halos to some extent at the masses of interest \citep[e.g.,][]{Schaller15, Xu17, Peirani17}.}

We adopt \editth{a composite mass distribution consisting of the contracted dark matter distribution and the stellar distribution} to model the stellar kinematics of the lens galaxies in our sample. For computational simplicity, we use spherical cases of these profiles. This assumption is sufficient for the quality of the measured velocity dispersions \citep{Sonnenfeld12}. The line-of-sight velocity dispersion can be expressed by solving the spherical Jeans equation as
\begin{equation}
	\sigma_{\rm los}^2 (R) = \frac{2G}{I(R)} \int_R^\infty \mathcal{K}_{\beta}\left(\frac{r}{R}\right) \frac{l(r) M(r)}{r}\ \mathrm{d}r,
\end{equation}
where $G$ is the gravitational constant, $I(R)$ is surface brightness distribution, $M(R)$ is the 3D enclosed mass, $l(r)$ is the 3D luminosity density, and $\mathcal{K}_{\beta}$ is a function that depends on the parametrization of the anisotropy parameter $\beta$ \citep{Mamon05}. We adopt the Osipkov--Merritt anisotropy profile given by
\begin{equation}
	\beta(r) \equiv 1 - \frac{\sigma_{\rm t}^2}{\sigma_{\rm r}^2} = \frac{r^2}{r^2 + a_{\rm ani}^2 R_{\rm eff}^2}	,
\end{equation}
where \editfv{$\sigma_{\rm r}^2$ is the radial velocity dispersion, $\sigma_{\rm t}^2$ is the tangential velocity dispersion, and} $a_{\rm ani}$ is the anisotropy scaling factor \citep{Osipkov79, Merritt85b, Merritt85}. For this parametrization, the function $\mathcal{K}_{\beta}$ has the form
\begin{equation}
\begin{split}
	\mathcal{K}_{\beta} (u) &= \frac{u_{\rm ani}^2 + 1/2}{(u_{\rm ani}+1)^{3/2}}	\left( \frac{u^2+u_{\rm ani}^2}{u} \right) \tan^{-1} \left( \sqrt{\frac{u^2-1}{u^2_{\rm ani}+1}} \right)  \\
	&\qquad\qquad\qquad - \frac{1/2}{u_{\rm ani}^2 + 1} \sqrt{1 - \frac{1}{u^2}},
\end{split}	
\end{equation}
where $u_{\rm ani} \equiv a_{\rm ani} R_{\rm eff}/R$ \citep{Mamon05}. \editref{The observed SDSS velocity dispersion is averaged over the aperture of 1.5 arcsec radius as
\begin{equation}
	\sigma_{\rm ap}^2 = \frac{\int_{\rm ap} \left[I(R)\sigma_{\rm los}^2(R) \circ \mathcal{S}\right] \ R\mathrm{d}R \mathrm{d}\theta}{\int_{\rm ap} \left[ I(R) \circ \mathcal{S}\right] \ R\mathrm{d}R \mathrm{d}\theta},
\end{equation}
where $\circ\mathcal{S}$ denotes convolution with the seeing.
}

\editth{We adopt the NFW profile for the initial distribution of the dark matter, which is given by
\begin{equation}
	\rho_{\rm NFW}(r)  \equiv \frac{\rho^{\rm NFW}_{\rm s}}{\left(r/r^{\rm NFW}_{\rm s}\right)\left(1 + r/r^{\rm NFW}_{\rm s}\right)^2},
\end{equation}
where $\rho_{\rm s}^{\rm NFW}$ is the density normalization and $r_{\rm s}^{\rm NFW}$ is the scale radius. For the stellar mass distribution we first adopt a constant mass-to-light ratio ($M/L$) in Section \ref{sec:constant_ml}, and then an $M/L$ gradient in Section \ref{sec:ml_gradient}. We adopt a double S\'ersic profile for the stellar light distribution similar to our lens models. To ensure robustness of the light profile fits, we fit the double S\'ersic profile from 20$\times$20 arcsec$^2$ cut-outs around the lens galaxies, which sufficiently contain \editref{the galaxy light that is above the background level}. Before fitting the light profiles, we subtract the lensed arcs using our best fit lens models. We only obtain the best fit double S\'ersic profile and adopt an uncertainty on the fit equivalent to 2 per cent uncertainty for the effective radius. This 2 per cent uncertainty is a conservative estimate, as the 2 per cent corresponds to 95 percentile of the uncertainty distribution of the effective radii in our lens models. We use the concentric Gaussian decomposition method from \citet{Shajib19b} to deproject the 2D stellar light or mass distribution into the corresponding 3D distribution. We take 30 Gaussian components with their standard deviations logarithmically spaced between 0.001 and 30 arcsec. This decomposition approximates the 2D stellar mass profile shapes within $\sim$0.2 per cent accuracy between 0.005 and 6 arcsec.}

\subsubsection{Constant $M/L$ for stellar mass distribution} \label{sec:constant_ml}

We first constrain the model parameters $w \equiv \{\nu,\ a_{\rm ani},\ f_{\rm gal},\ r_{\rm s}^{\rm NFW},\ \rho_{\rm s}^{\rm NFW} \}$ for individual lens systems. Here, $r_{\rm s}^{\rm NFW}$ is the scale radius of the NFW profile and $\rho_{\rm s}^{\rm NFW}$ is the normalization. 
\editth{We impose a theoretical prior on the $M_{200}$--$c_{200}$ relation corresponding to $\langle z_{\rm d} \rangle = 0.20$ from \citet{Diemer19} for the initial NFW halo.} \editth{To incorporate the lensing constraints into this joint \editfv{lensing--dynamics} analysis,} we \editf{fold in the posterior distributions of Einstein radius $R_{\rm E}$ and the \editth{quantity $R_{\rm E} \alpha^{\prime\prime}_{\rm E} / (1 - \kappa_{\rm E})$} from our lens models in Section \ref{sec:slope_dist}}. \editth{Here, $\alpha^{\prime\prime}_{\rm E}$ is the double derivative of the deflection angle at $R_{\rm E}$ and the $\kappa_{\rm E}$ is the convergence at $R_{\rm E}$. The term $R_{\rm E} \alpha^{\prime\prime}_{\rm E} / (1 - \kappa_{\rm E})$ is the mass-model-independent observable quantity from imaging data \citep{Kochanek20}.} \editth{We also use the weak lensing measurements of the reduced shear, although weak lensing \editfv{do not provide tight constraints due} to large uncertainty in the measured shear of individual systems.} We take uniform priors on $\nu \sim \mathcal{U}(-0.5,\ 1)$, $a_{\rm ani} \sim \mathcal{U}(0.1,\ 5)$, and $f_{\rm gal} \sim \mathcal{U}(0,\ 0.25)$. \editfr{Additionally we impose a prior on $\log_{10} M_{200}$ from \citet{Sonnenfeld18b} that depends on the measured stellar mass from the stellar population synthesis method assuming Chabrier IMF. We obtain the Chabrier IMF based stellar masses for the galaxies in our sample from \citet{Auger09}. The form of the prior is
\begin{equation}
	\begin{aligned}
	&p(\log_{10} M_{200} \mid \log_{10} M_{\star}^{\rm Chab}) \\\
	&\qquad \equiv \mathcal{N} \left(\mu_{\rm h} + \beta_{\rm h} \left[\log_{10} M_{\star}^{\rm Chab} - 11.3\right],\ \sigma_{\rm h}^2 \right).
	\end{aligned}
\end{equation}
The parameters in this prior are obtained by \citet{Sonnenfeld18b} by applying the SLACS selection function on SDSS galaxies that have HSC weak lensing data and assuming a model with $M/L$ gradient for the stellar mass and an NFW profile for the dark matter. We take the uncertainty on these parameters to match with the dispersion observed in the SLACS galaxies by \citet{Sonnenfeld18b} for the same mass model to have $\mu_{\rm h} = 13.03 \pm 0.25$, $\beta_{\rm h} = 1.42 \pm 0.92$, and $\sigma_{\rm h} = 0.18 \pm 0.16$.
}


We then infer the distribution of population-level parameters $\tau$ for our lens sample using hierarchical Bayesian inference. According to the Bayes' theorem, we can express the posterior of the population-level parameters as
\begin{equation} \label{eq:hier_integral}
\begin{aligned}
	p(\tau \mid D) &\propto p(D \mid \tau) \ p(\tau) \\
	&\propto p(\tau) \ \prod_{i}^{N} \int \mathrm{d}w_{i} \ p(D_i \mid w_i) \ p(w_i \mid \tau) \\
	&\propto p(\tau) \ \prod_{i}^{N} \int \mathrm{d}w_{i} \ \frac{p(w_i \mid D_i)}{p(w_i)} \ p(w_i \mid \tau),
\end{aligned}
\end{equation}
Here, \editfv{$D \equiv \left\{ D_{i} \right\}$ refers to the data set containing all the data $D_i$ for individual lenses}, $p(D \mid \tau)$ is the likelihood term on the population-level parameters $\tau$, $p(\tau)$ is the prior, $p(w_i \mid \tau)$ governs the distribution of individual lens parameters given the population-level parameters, \editfr{$p(w_i \mid D_i)$ is the posterior of $w_i$ for an individual system given its associated data $D_i$}, \editfv{and $p(w)$ is the prior for the parameter set $w$ as given above.} \editref{The likelihood function $p(D_i \mid w_i)$ for the $i$th lens system has the form
\begin{equation}
	\begin{aligned}
	p(D_i \mid w_i) \equiv &\mathcal{N}(R_{\rm E, i},\ \sigma_{R_{\rm E}, i}^2) \times \mathcal{N}(R_{\rm E} \alpha^{\prime\prime}_{\rm E} / (1 - \kappa_{\rm E}),\ \sigma^2_{[R_{\rm E} \alpha^{\prime\prime}_{\rm E} / (1 - \kappa_{\rm E})]}) \\ &\times \mathcal{N}(\sigma_{\rm ap},\ \sigma_{\sigma_{\rm ap}}^2) \times \prod_j \mathcal{N}(\tilde{\gamma}_{{\rm shear}, ij},\ \sigma^2_{\tilde{\gamma}_{{\rm shear}, ij}}),
	\end{aligned}
\end{equation}
where $\sigma_{\rm ap}$ is the observed velocity dispersion, and $\tilde{\gamma}_{{\rm t}, ij}$ is the reduced weak-lensing shear in the $j$th radial bin.} We assume that $a_{\rm ani}$, $\nu$, and $f_{\rm gal}$ are Gaussian-distributed for our lens sample \editsx{with the Gaussians truncated where $p(w) = 0$}. Thus, we take $\tau \equiv \{\mu_{\nu},\ \sigma_{\nu},\  \mu_{a_{\rm ani}},\ \sigma_{a_{\rm ani}},\ \mu_{f_{\rm gal}},\ \sigma_{f_{\rm gal}} \}$ as the population-level parameters.  Here, $\mu$ and $\sigma$ refer to the mean and standard deviation of the population-level Gaussian distribution for the corresponding parameter. We take uniform priors $\mu_{\nu} \sim \mathcal{U}(-1, 1)$, $\mu_{a_{\rm ani}} \sim \mathcal{U}(0.5, 5)$, and $\mu_{f_{\rm gal}} \sim \mathcal{U}(0, 0.25)$. We take the same uniform prior $\mathcal{U}(-5, 1)$ for $\log_{10} \sigma_{\nu}$, $\log_{10} \sigma_{a_{\rm ani}}$, and $\log_{10} \sigma_{f_{\rm gal}}$. We perform the integral in equation (\ref{eq:hier_integral}) using Monte Carlo integration. 
\editet{In the Monte Carlo integration, we sample from $p(w_i \mid \tau)$ and sum over $p(w_i \mid D_i) / p(w_i)$. We approximate $p(w_i \mid D_i)$ with a Gaussian mixture model (GMM) of the corresponding samples drawn using MCMC. Before fitting the GMM, we smooth the distributions of the MCMC samples using the kernel density estimation (KDE) method with Silverman's rule for the bandwidth \citep{Silverman86}. The number of components in the GMM for each individual lens posterior $p(w_i \mid D_i)$ is selected using Bayesian information criterion (BIC) and it is typically within 5--10. \editref{We check the robustness of our hierarchical inference framework with simulated distributions of $p(w_i \mid D_i)$ and find that the correct population-level parameter posterior $p(\tau \mid D)$ is recovered.}
}

\editf{\editth{We illustrate the posterior distributions of the population-level parameters in Figure \ref{fig:hierarchical_dist}, and tabulate their point estimates in Table \ref{tab:hier_params}.}  \editfr{In Figure \ref{fig:hierarchical_dist_interpretation}, we highlight some of the important results from Figure \ref{fig:hierarchical_dist} along with their interpretations for galaxy properties.} We infer the mean halo response parameter \editet{$\mu_{\nu} = -0.06\pm0.04$} and intrinsic scatter \editet{$\sigma_{\nu} \leq 0.092$ (95 per cent upper limit)}, which is consistent \editet{within 1.5$\sigma$} with no contraction on average for our lens sample.

\editth{We compute the dark matter fraction $f_{\rm dm}$ within $R_{\rm eff}/2$ directly from the decomposed dark and stellar mass distributions. We show the $\gamma$--$f_{\rm dm}$ distribution in Figure \ref{fig:f_dm_w_kin}. We find moderate anticorrelation between $\gamma$ and $f_{\rm dm}$ with the biweight mid-correlation \editfv{$r=-0.39 \pm 0.15$}. The anticorrelation is expected as a shallower total mass profile slope will require higher contribution from the dark matter to lower the total mass profile's slope. The distribution of $f_{\rm dm}$ in our sample has a mean of \editfv{$\langle f_{\rm dm} \rangle = 0.27\pm0.02$} and a scatter of \editfv{$0.14 \pm 0.01$}.}


\begin{figure*}
	\includegraphics[width=1.\textwidth]{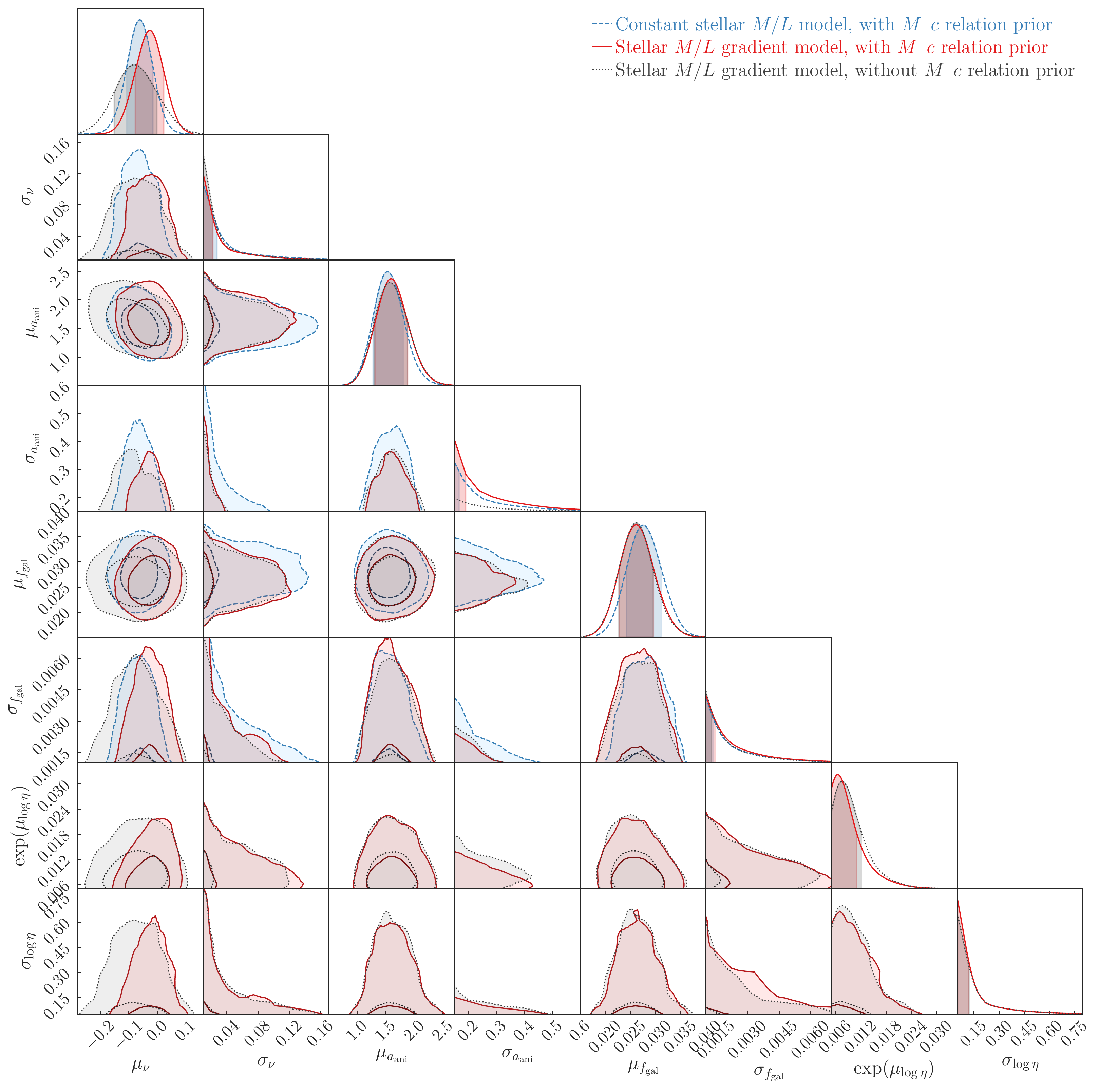}
	\caption{ \label{fig:hierarchical_dist}
		Distribution of population-level parameters assuming Gaussian distribution for the halo response parameter $\nu$, the anisotropy scaling parameter $a_{\rm ani}$, \editth{fraction of the total mass that cools to form stars $f_{\rm gal}$, and $\log \eta$ where $\eta$ is the $M/L$ gradient's exponent.} \editet{The three model setups illustrated are a constant stellar $M/L$ model with the $M$--$c$ relation prior (blue dashed line), a stellar $M/L$ gradient model with the $M$--$c$ relation prior (red solid line), and a stellar $M/L$ gradient model without the $M$--$c$ relation prior (dotted grey line).} The darker and lighter regions represent 68 per cent and 95 per cent credible regions, respectively. The 2D contours and 1D distributions are smoothed with a Gaussian kernel of \editet{1.5$\sigma$} bandwidth. The point estimate of these parameters are given in Table \ref{tab:hier_params}. \editfr{Figure \ref{fig:hierarchical_dist_interpretation} highlights the central parameters in our model along with their interpretations.}}
\end{figure*}

\begin{figure*}
	\includegraphics[width=1.\textwidth]{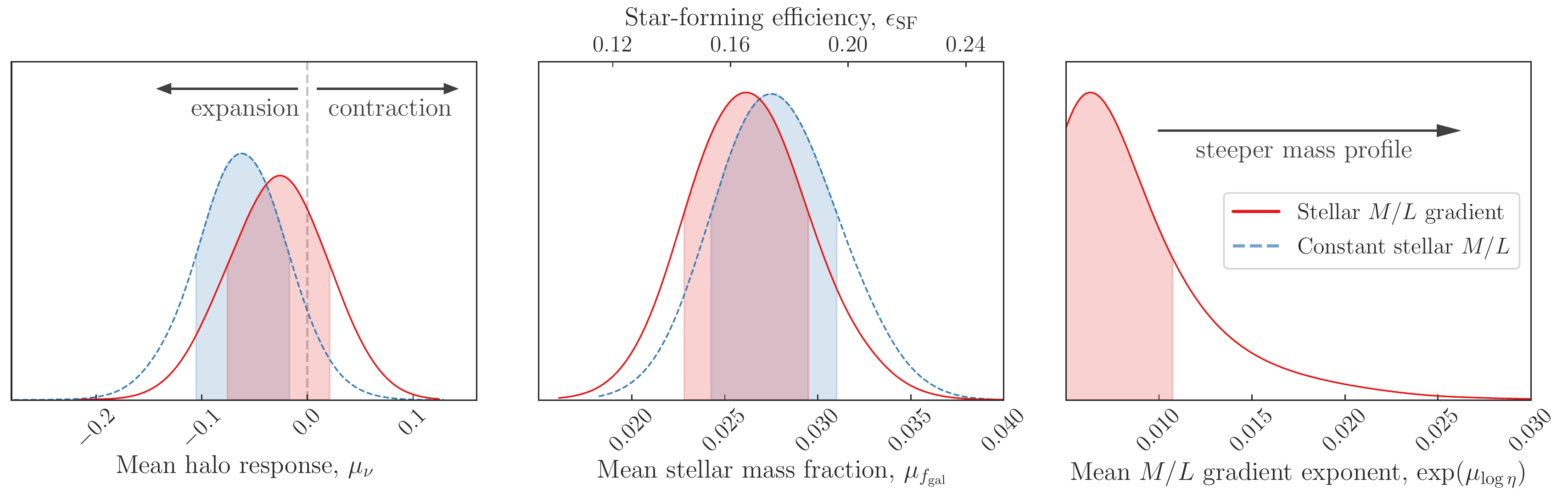}
	\caption{ \label{fig:hierarchical_dist_interpretation}
		\editfr{Interpretation of 1D marginalized distributions of some central model parameters for our galaxy sample. \textbf{Left-hand panel:} distributions of the sample mean $\mu_{\nu}$ of the halo response parameter $\nu$ for stellar $M/L$ gradient (red solid line) and constant stellar $M/L$ (blue dashed line) models. Both models are consistent with zero contraction marked by the vertical grey dashed line. The full contraction scenarios in \citet{Blumenthal86} model ($\nu =1$), and the simulations of \citet[][$\nu = 0.8$]{Gnedin04} and \citet[][$\nu \sim 0.4$]{Abadi10} are ruled out. \textbf{Middle panel:} distributions of the mean fraction of star-forming baryons $\mu_{f_{\rm gal}}$ that has cooled from the initial distribution. We show along the top axis the corresponding star-forming efficiency  $\epsilon_{\rm SF} \equiv f_{\rm gal} \Omega_{\rm m}/\Omega_{\rm b}$, where $\Omega_{\rm b}/\Omega_{\rm m}$ is adopted from \citet{PlanckCollaboration18}. The $\mu_{f_{\rm gal}}$ distributions peaks at $\epsilon_{\rm SF} \sim 0.17$, which is consistent with moderate to strong feedback mechanisms on the star formation \citep{Hopkins14}. \textbf{Right-hand panel:} The distribution of mean $M/L$-gradient-exponent in our $M/L$ gradient model. The distribution favours very small gradient, \editfv{with the 95 per cent upper limit at 0.02}.}}
\end{figure*}

\renewcommand{\arraystretch}{1.5}
\begin{table*}
\caption{\label{tab:hier_params}
		\editth{1D marginalized distributions of the population-level parameters from Bayesian hierarchical inference. The corner plot for these parameters are illustrated in Figure \ref{fig:hierarchical_dist}.} \editfv{The baseline model settings are: $M$--$c$ relation prior from \citet{Diemer19}. Jeffrey's prior for $M/L$-gradient exponent $\eta$, and $H_0 = 70$ km s$^{-1}$ Mpc$^{-1}$. ``Other settings'' column indicates which setting from these baseline settings is varied. The columns \editet{for $\sigma_{\nu}$, $\sigma_{a_{\rm ani}}$, $\sigma_{f_{\rm gal}}$,} $\exp(\mu_{\log \eta})$\editet{, and $\sigma_{\log \eta}$ provide} the 95 per cent upper limit.}
	}
\resizebox{\textwidth}{!}{\begin{threeparttable}

	\begin{tabular}{llcccccccc}
	\hline
	Stellar $M/L$ & Other settings & $\mu_{\nu}$ & $\sigma_{{\nu}}$ & $\mu_{a_{\rm ani}}$ & $\sigma_{a_{\rm ani}}$  & $\mu_{f_{\rm gal}}$ & $\sigma_{f_{\rm gal}}$ & $\exp(\mu_{\log \eta})$ & $\sigma_{\log \eta}$\\
\hline 

Constant $M/L$ & Baseline
& $-0.06_{-0.04 }^{ +0.04 }$
& $\leq$0.092
& $1.54_{-0.22 }^{ +0.25 }$
& $\leq$0.26
& $0.028_{-0.003 }^{ +0.003 }$
& $\leq$0.0040
& -- & -- \\
$M/L$ gradient & Baseline
& $-0.03_{-0.05 }^{ +0.04 }$
& $\leq$0.074
& $1.60_{-0.25 }^{ +0.27 }$
& $\leq$0.22
& $0.026_{-0.003 }^{ +0.003 }$
& $\leq$0.0041
& $\leq$0.017
& $\leq$0.33
\\
$M/L$ gradient & No $M$--$c$ prior
& $-0.08_{-0.06 }^{ +0.07 }$
& $\leq$0.069
& $1.60_{-0.25 }^{ +0.28 }$
& $\leq$0.20
& $0.026_{-0.003 }^{ +0.003 }$
& $\leq$0.0036
& $\leq$0.017
& $\leq$0.36
\\
$M/L$ gradient & $\eta$ prior $\sim \mathcal{U}(0,\ 1)$ 
& $0.02_{-0.04 }^{ +0.04 }$
& $\leq$0.054
& $1.52_{-0.23 }^{ +0.26 }$
& $\leq$0.25
& $0.025_{-0.003 }^{ +0.003 }$
& $\leq$0.0036
& ($\leq$0.061)$^{\rm a}$
& ($\leq$0.015)$^{\rm a}$
\\

\hline
	\end{tabular}
	\begin{tablenotes}
	     \item[$^{\rm a}$] For the model with uniform prior on $\eta$, the column $\exp(\mu_{\log \eta})$ gives the value for $\mu_{\eta}$ and the column $\sigma_{\log \eta}$ gives the value for $\sigma_{\eta}$.
	\end{tablenotes}
	\end{threeparttable}
	}
\end{table*}

\begin{figure}
	\includegraphics[width=\columnwidth]{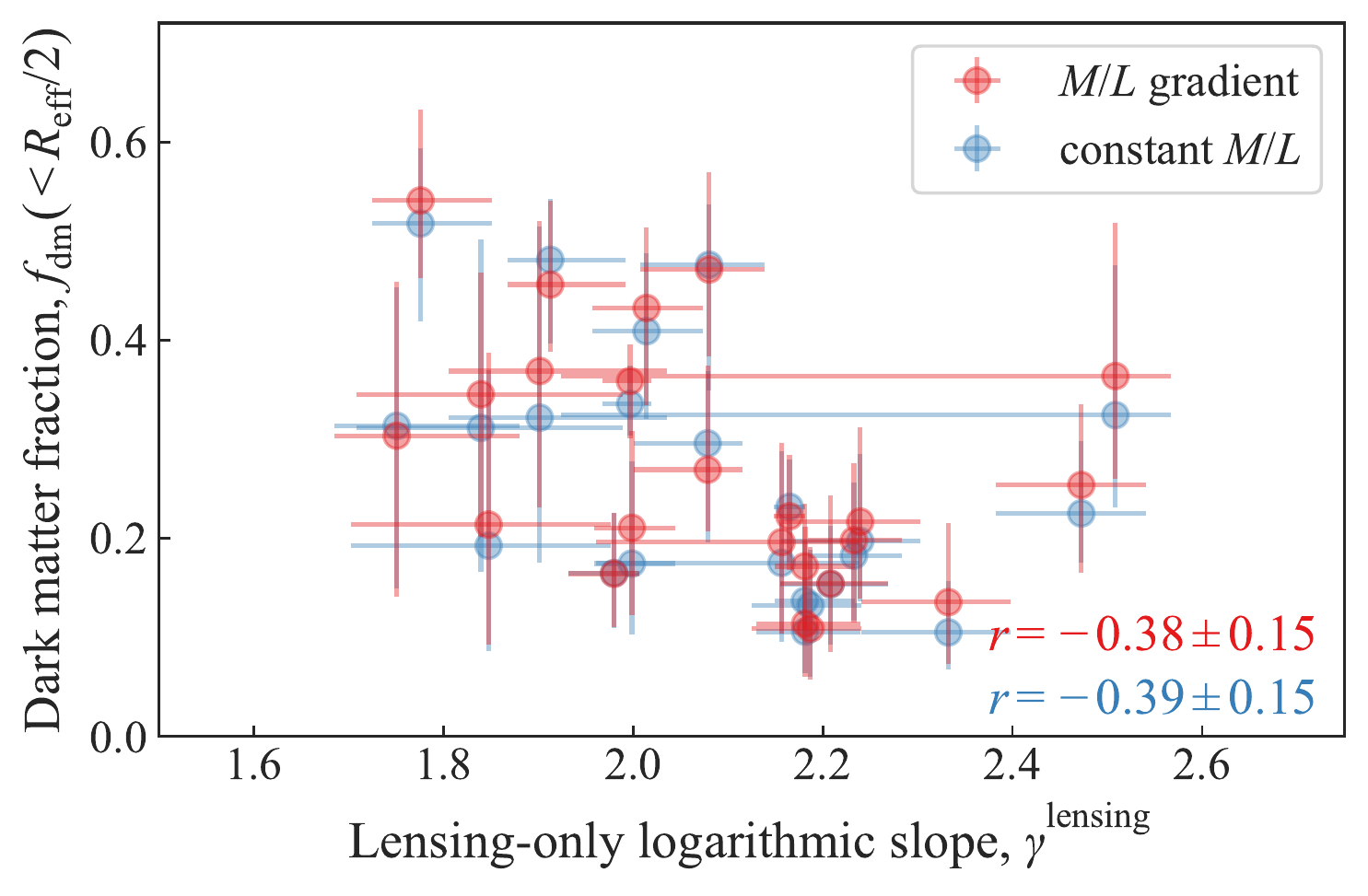}
	\caption{ \label{fig:f_dm_w_kin}
	Distribution of the local logarithmic slope $\gamma$ and the dark matter fraction $f_{\rm dm}$ within half effective radius. The blue circles are from the constant $M/L$ model for the stellar mass distribution. \editfr{$f_{\rm dm}$ is moderately anticorrelated with $\gamma$, as higher stellar contribution is needed to increase the logarithmic slope, which brings the dark matter fraction $f_{\rm dm}$ down.}
	}
\end{figure}

\begin{figure*}
	\includegraphics[width=.9\textwidth]{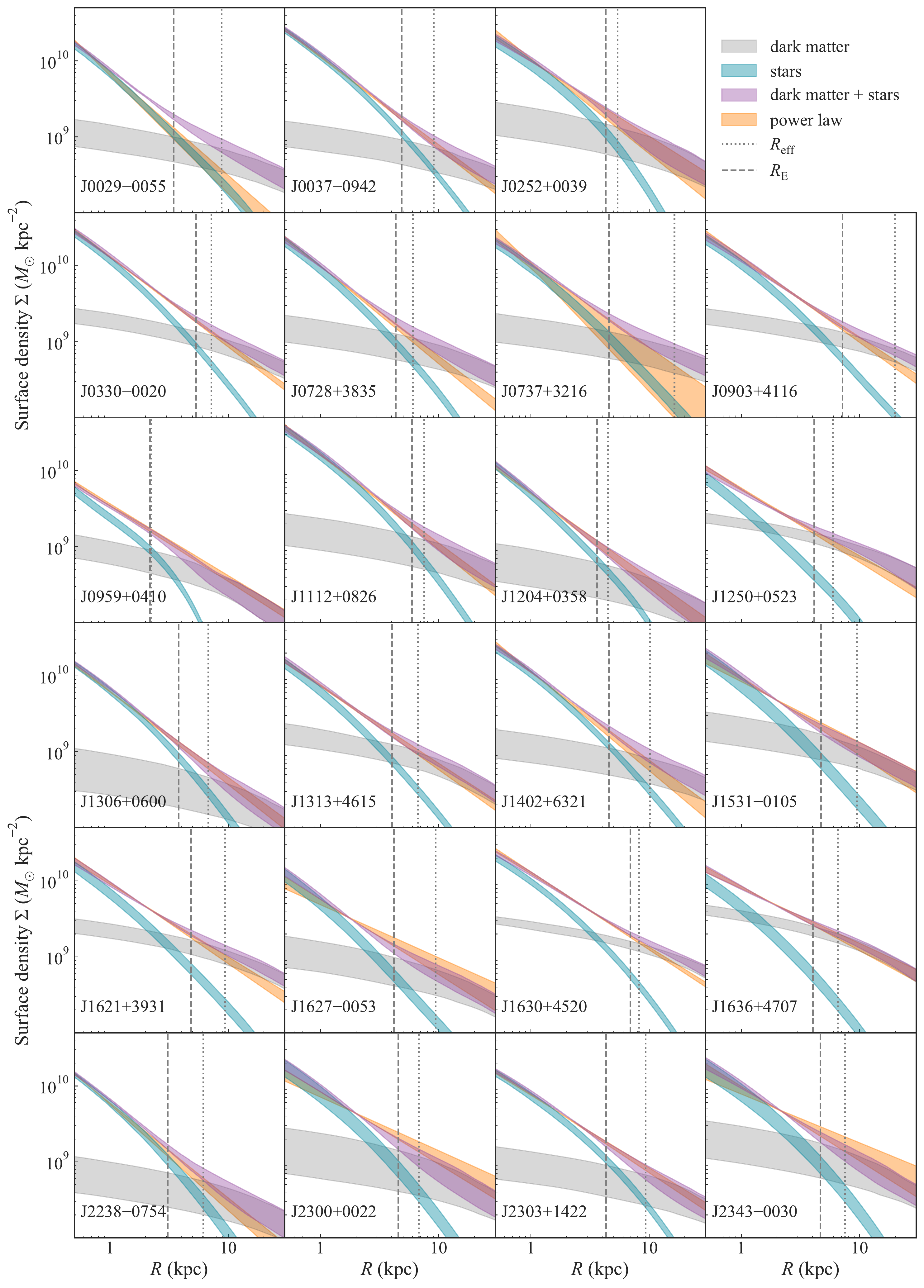}
	\caption{ \label{fig:profile_shape}
		Shape of the surface density profile for total and component mass distributions. The purple shaded region shows the 1$\sigma$ credible region for the total mass distribution from joint strong-lensing--kinematics--weak lensing constraints. The grey and \editfr{teal blue} shaded regions show the dark and stellar component distributions, respectively. The power-law profile from our lens models are shown with the orange shaded region. The Einstein and the effective radii for each lens system are marked with the dashed grey line and the dotted grey line, respectively.}
\end{figure*}

\begin{figure*}
	\includegraphics[width=.9\textwidth]{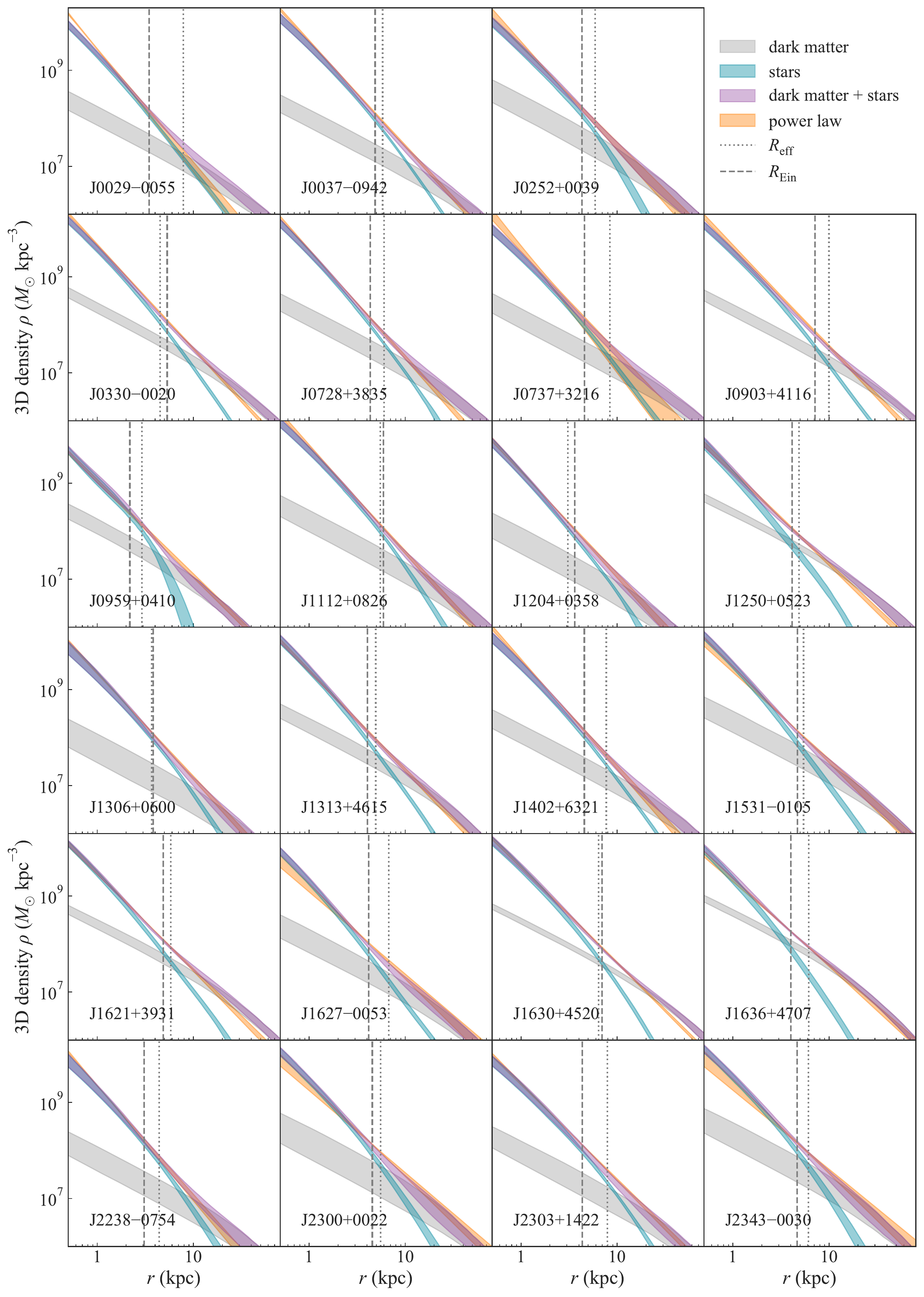}
	\caption{ \label{fig:profile_shape3d}
		\editfr{Shape of the 3D density profile for total and component mass distributions from spherical deprojection from the surface density profiles shown in Figure \ref{fig:profile_shape}. The purple shaded region shows the 1$\sigma$ credible region for the total mass distribution from joint strong-lensing--kinematics--weak lensing constraints. The grey and \editfr{teal blue} shaded regions show the dark and stellar component distributions, respectively. The power-law profile from our lens models are shown with the orange shaded region. The Einstein and the effective radii for each lens system are marked with the dashed grey line and the dotted grey line, respectively.}}
\end{figure*}

\begin{figure*}
	\includegraphics[width=\textwidth]{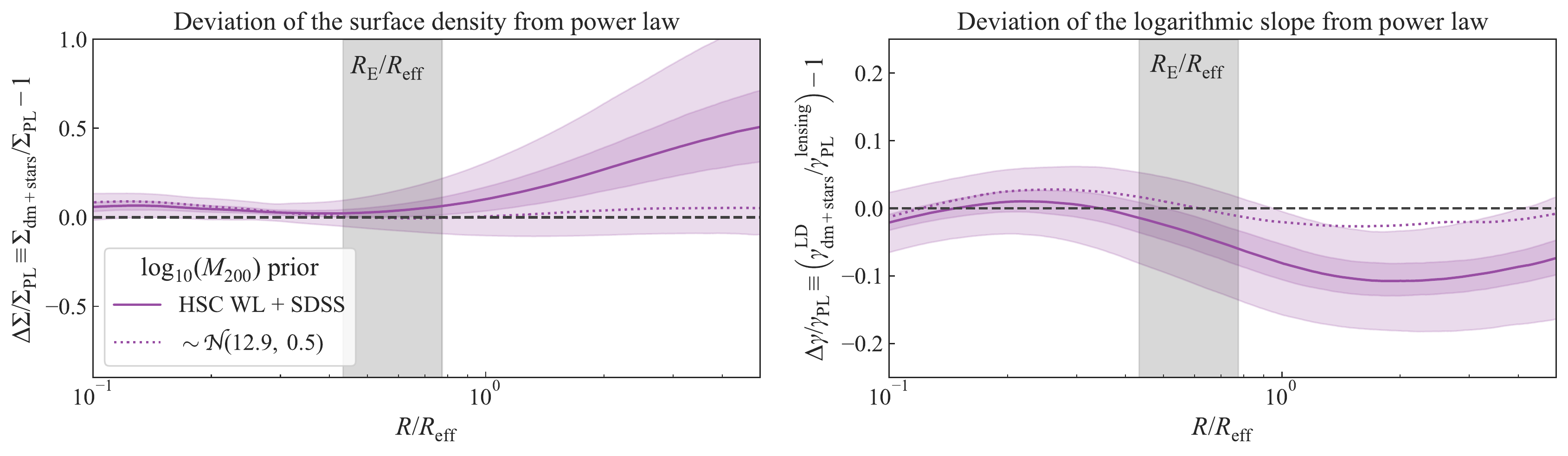}
	\caption{ \label{fig:deviation_from_power_law}
		\editfr{Deviation of the surface density profile (left) and the local logarithmic slope (right) in the NFW+stars mass distribution from the power-law mass distribution. The NFW+stars mass distribution is constrained from joint \editref{strong-lensing--dynamics--weak-lensing} analysis, whereas the power-law mass distribution is constrained from lensing-only data. The solid purple line shows the median deviation for our adopted prior based on HSC weak lensing measurements of SDSS galaxies. The darker and lighter purple shaded regions illustrate the 1$\sigma$ credible interval of the median and the median absolute deviation, respectively. The grey dashed line represent \editfv{no} deviation from the power-law. The dotted purple line correspond to the median deviation if a prior with a lower mean halo mass is adopted. The vertical grey shaded region shows the 1$\sigma$ distribution of $R_{\rm E}/R_{\rm eff}$ for our sample. The deviation of the NFW+stars  mass distributions from than the power-law estimate at the Einstein radius has a median of $\sim$5 per cent and median absolute deviation of $\sim$7 per cent. The deviation of the logarithmic slopes at the Einstein radius has a median of $\sim$2 per cent and median absolute deviation of $\sim$2 per cent.
		}}
\end{figure*}

\subsubsection{$M/L$ gradient for stellar mass distribution} \label{sec:ml_gradient}

\editth{Next, we incorporate a $M/L$ gradient for the stellar mass distribution. We parameterize a power-law $M/L$ gradient as 
\begin{equation}
	\Upsilon (R) = \Upsilon_{\rm e} \left( \frac{R}{R_{\rm eff}}\right)^{-\eta},
\end{equation}
where $\Upsilon_{\rm e}$ is the $M/L$ at $R_{\rm eff}$. \editfv{However}, this parametrization diverges for $\eta > 0$ as $R \to 0$, the total stellar mass computed using the concentric Gaussian decomposition method stays finite. Moreover, the observed lensing properties and the kinematic properties are mostly sensitive to the mass distribution near the Einstein radius and the effective radius, respectively. As a result, the deviation from the exact relation near the centre ($\lesssim$0.01 arcsec) in our Gaussian-decomposed approximation does not noticeably impact our analysis.}

\editth{We perform the same analysis from Section \ref{sec:constant_ml}, but with the $M/L$ gradient implemented in the stellar mass distribution. As a result, the model parameters for individual lenses are extended to $w \equiv \{\nu,\ a_{\rm ani},\ f_{\rm gal},\ r_{\rm s}^{\rm NFW},\ \rho_{\rm s}^{\rm NFW},\ \eta \}$. We only allow positive values for $\eta$ as previous observations suggest that the $M/L$ is higher at the centre than outer region of elliptical galaxies \citep[e.g.,][]{MartinNavarro15, vanDokkum17}. As we want to adopt an uninformative prior on the order of magnitude of $\eta$, we take a Jeffrey's prior on $\eta$ as $p(\eta) \propto 1/\eta$, which is equivalent to a uniform prior on $\log\eta$. We set the bounds of the uniform prior on $\log \eta$ as \editet{$p(\log \eta) \sim \mathcal{U}(\log 10^{-3},\ 0)$}. Furthermore, the population-level parameters for the Bayesian hierarchical inference are also extended to $\tau \equiv \{\mu_{\nu},\ \sigma_{\nu},\  \mu_{a_{\rm ani}},\ \sigma_{a_{\rm ani}},\ \mu_{f_{\rm gal}},\ \sigma_{f_{\rm gal}},\ \mu_{\log \eta},\ \sigma_{\log \eta} \}$.
}

\editth{We infer the \editet{95 per cent upper limit} of $\exp(\mu_{\log \eta})$ distribution to be 0.02. The mean halo response parameter for the sample is \editet{$\mu_{\nu} = -0.03^{+0.04}_{-0.05}$} with \editfv{intrinsic} scatter \editet{$\sigma_{\nu} \leq 0.074$ (95 per cent upper limit)}, which is consistent with no contraction in the NFW halo (Figure \ref{fig:hierarchical_dist}). We show the distribution of the inferred $f_{\rm dm}$ and the logarithmic slope $\gamma$ in Figure \ref{fig:f_dm_w_kin}. The distribution of $f_{\rm dm}$ in our sample for the $M/L$ gradient model has a mean \editfv{$ \langle f_{\rm dm} \rangle = 0.28 \pm 0.02$} and an \editfv{intrinsic} scatter of \editfv{$0.15 \pm 0.02$}.}

\editth{To check for the impact of the $M$--$c$ relation prior in this analysis, we perform the joint \editfv{lensing--dynamics} analysis without the $M$--$c$ relation prior. We find that the uncertainties on the model parameters \editfv{for each lens system} expectedly increase without this prior, and the posteriors from with or without the prior are consistent within $1\sigma$ \editfv{(Figure \ref{fig:hierarchical_dist}, Table \ref{tab:hier_params})}. 
}}

\editfv{We also check for the impact of our fiducial cosmology, in particular the Hubble constant, on our inference. We perform the joint lensing--dynamics analysis with $H_0 = 67$ km s$^{-1}$ Mpc$^{-1}$ and with $H_0 = 74$ km s$^{-1}$ Mpc$^{-1}$. We find all the posterior distributions to be consistent within 1$\sigma$. 
Therefore, our inference is not sensitive to the choice of the Hubble constant.}

\editth{We show the surface density profiles and the deprojected 3D density profiles of the dark matter and stellar components of the total mass distribution for all the lens galaxies in Figures \ref{fig:profile_shape} and \ref{fig:profile_shape3d}, respectively.} \editfr{In Figure \ref{fig:deviation_from_power_law}, we illustrate the median deviation of the NFW+stars model from the power law in the total surface density profile and the local logarithmic slope. The total density profile combining dark matter and stars deviates \editfv{higher} by $\sim$5 per cent (median value) at the Einstein radius with a median absolute deviation of $\sim$7 per cent within our sample. Here, the median absolute deviation is a proxy for the intrinsic scatter. The local slope of the density profile at the Einstein radius is shallower by $\sim$2 per cent (median value) with median absolute deviation of $\sim$2 per cent. These deviations are sensitive to our adopted prior on $M_{200}$. If we choose a prior with lower mean halo mass such as $\log_{10} M_{200} \sim \mathcal{N} (12.9, 0.5)$ -- which corresponds to the BOSS constant-mass (CMASS) galaxies with mean redshift $z\approx0.6$ \citep{Sonnenfeld19b} -- the resultant \editfv{profiles would be within 0.8 per cent of the power laws at the Einstein radius} (Figure \ref{fig:deviation_from_power_law}).}

\section{Discussion and Comparison with previous studies} \label{sec:discussion}

\editfr{In this section, we compare our results with previous studies based on observations and simulations, and interpret them in the context of elliptical galaxy evolution. First in Section \ref{sec:central_results}, we discuss the central results of this study on the elliptical galaxy structure.
We then place these results in the context of massive elliptical galaxy evolution in Section \ref{sec:evolution}.
We discuss our results on the $M/L$-gradient and its implication for the stellar IMF in Section \ref{sec:ml_gradient_discuss}. We discuss the implication of our results for time-delay cosmography in Section \ref{sec:time_delay_cosmography}. We state the limitations of this study in Section \ref{sec:limitations}.}

\subsection{Elliptical galaxy structure} \label{sec:central_results}

\editfr{
The main results of this study on the elliptical galaxy structure are: (i) the dark matter distribution in massive elliptical galaxies at $z\sim0.2$ are not contracted and close to an NFW profile on average (i.e., $\mu_{\nu}\sim 0$), and (ii) the total density profile combining the dark matter and stars is close to isothermal (i.e., $\langle \gamma \rangle \sim 2$). 
}

\subsubsection{Dark matter contraction}

\editfr{Our result on the dark matter contraction agrees with \citet{Dutton14}, who also find no contraction in SLACS galaxies using the lens models from \citet{Auger10b}. Similarly, \citet{Sonnenfeld15} and \citet{Newman15} find that the inner slope of the dark matter distribution is consistent with the NFW profile within the uncertainty for their samples of massive elliptical galaxy halos and group-scale halos, respectively. Therefore, these findings are consistent with our result. 
However, our result contradicts the report of steeper central slopes \editfv{(mean inner logarithmic slope $\langle {\beta} \rangle = 2.01^{+0.19}_{-0.22}$) than that in the NFW profile ($\beta = 1$)} by \citet{Oldham18}. }
\editth{\editfv{Interestingly}, \citet{Oldham18} find their slope distribution to be bimodal, and the steep value given above corresponds to the mode with the larger mean. \editfv{If \citet{Oldham18} adopt a unimodal slope distribution, then the average inner slope is consistent with the vanilla NFW profile within $0.5\sigma$ for the $M/L$ gradient model and within 1.3$\sigma$ for the constant $M/L$ model. Albeit, the unimodal distribution requires higher intrinsic scatter.} Moreover, \citet{Oldham18} note that their inferred halo mass is lower than the expectation of abundance matching studies due to the lack of a strong prior to constrain the halo scale size or its mass. Strong-lensing data is sensitive to the mass profile in the central region within the Einstein radius, which is typically $\sim$5--10 times smaller than the NFW scale radius. Thus, additional data or prior at scales larger than the NFW scale radius is necessary to robustly constrain the halo mass. We find that the degree of contraction depends on the halo mass prior in our analysis with a heavier prior on $M_{\rm 200}$ producing shallower inner slopes to fit the joint lensing--kinematics data. \editfv{The sample of \citet{Oldham18} has a similar stellar mass range as our sample and the difference between the mean redshifts of the samples $\Delta \langle z \rangle = 0.15$ does not leave enough room to expand the halos from $\beta\sim2$ to $\beta\sim1$ within $\sim$1.44 Gyr. Therefore, we conclude that the differences between our result and that from \citet{Oldham18} are largely caused by the difference in the adopted priors corresponding to the dark matter halo.}
}

\subsubsection{Slope of the total density profile}

\editfr{
Our lensing-only models provide $\langle \gamma^{\rm lensing} \rangle = 2.08 \pm 0.04$ with a scatter of 0.13$\pm$0.02. From the joint lensing--dynamics analysis, we find the total density profile is shallower by approximately 5 per cent, which brings the sample mean of the logarithmic slope at the Einstein radius closer to the isothermal case (Figure \ref{fig:deviation_from_power_law}). This near-isothermality of the total density profile agrees well with a multitude of pervious observations -- e.g., based on strong-lensing only or jointly based on lensing and dynamics: \citet{Treu04, Gavazzi07, Auger10b, Ritondale19}, and based on stellar dynamics: \citet{Thomas07, Tortora14, Bellstedt18}.
}

\begin{figure}
	\includegraphics[width=\columnwidth]{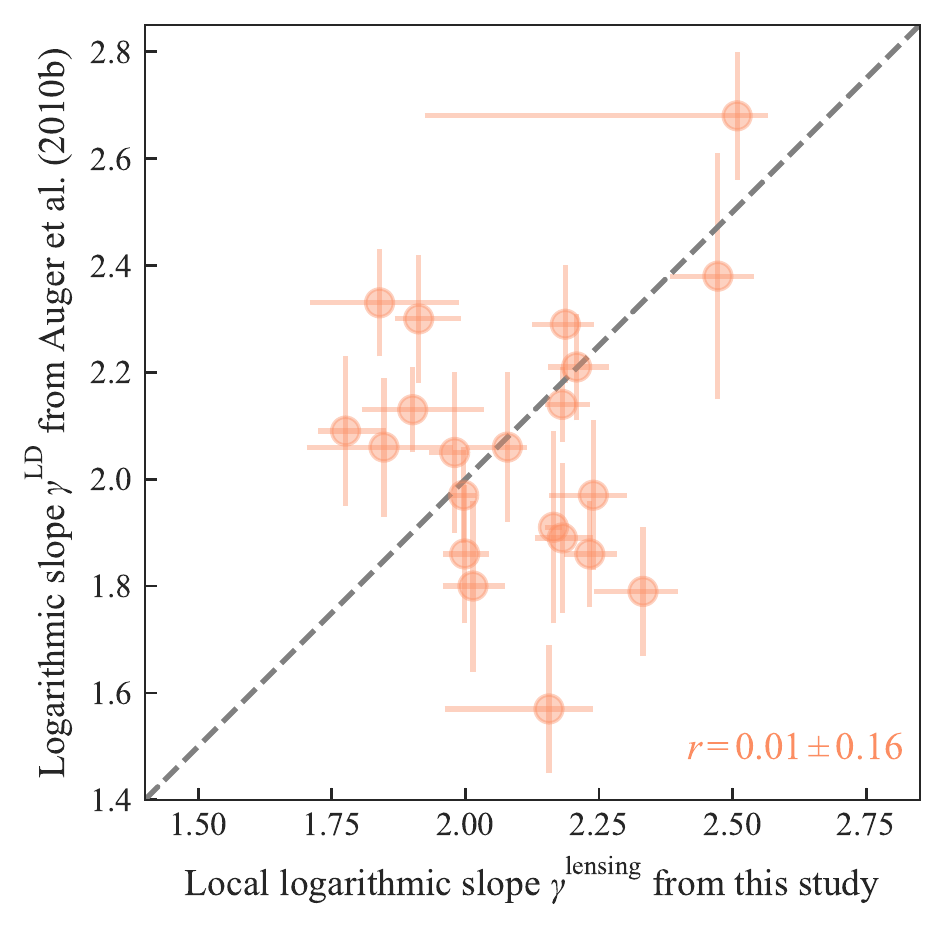}	
	\caption{ \label{fig:gamma_compare}
          Comparison between the estimated local logarithmic slopes $\gamma$ in this study, and \editth{the \editfr{radially averaged} logarithmic slope} estimated by the SLACS analysis \citep{Auger10b}. \editfr{Whereas \citet{Auger10b} constrain only the enclosed mass within the Einstein radius from the imaging data and then obtain the profile slope in combination with the stellar kinematics, we constrain both the enclosed mass and the local slope from the imaging data only.} This plot contains 21 systems that have $\gamma_{\rm SLACS}$ measured by \citet{Auger10b}. The dashed grey line illustrates a hypothetical perfect correlation to guide the visualization. We find no correlation between the two measurements with biweight mid-correlation $r=0.01 \pm 0.16$. \editth{Given the measurement uncertainties and an additional 12 per cent uncertainty on $\gamma_{\rm SLACS}$ -- which is equivalent to 6 per cent unaccounted systematic uncertainty in the measured kinematics \citep{Birrer20} -- \ttedit{we cannot conclude based on this comparison whether the expected correlation for a global power law is present under the noise, or not}.}}
\end{figure}

In Figure \ref{fig:gamma_compare}, we compare the distribution of the estimated logarithmic slopes $\gamma$ constrained from the imaging data only in this study with those estimated by the SLACS analysis from combining stellar kinematics with the imaging data. We find no correlation between the estimated $\gamma$ distributions from the two analyses with biweight mid-correlation $r = 0.03 \pm 0.17$. However, the SLACS distribution has a mean of $\langle \gamma^{\rm LD} \rangle = 2.078 \pm 0.027$ and intrinsic scatter $0.16 \pm 0.02$ \citep{Auger10b}. These values are consistent with our results within 1$\sigma$ confidence level. \editth{For the 21 systems that have measured $\gamma$ in \citet{Auger09}, we find $\Delta_{\gamma}^2 \equiv \sum_i^{N=21}(\gamma^{\rm lensing}_{i, \rm this\ study} - \gamma^{\rm LD}_{i, \rm Auger+(10b)})^2/\sigma_{i, \rm total}^2 = 21.76$. The p-value assuming a $\chi^2$-distribution for $ \Delta_{\gamma}^2$ with 21 degrees of freedom is 0.41. \editfr{We can see in Figure \ref{fig:profile_shape} that the two-component mass profile from lensing--dynamics can deviate from the lensing-only inference of the power-law profile towards either direction. However, the sample mean of the such deviations is smaller than $\sim$5 per cent near the Einstein radius, which explains the good agreement for $\langle \gamma \rangle$ between this study and \citet{Auger10b}.  Thus, a correlation between the lensing-only local slope and the lensing--dynamics global slope is expected. However, \editfv{such correlation cannot be detected within the noise}. 
}
}

\editref{We assume Osipkov--Merritt anisotropy profile as this profile follows the observation that the anisotropy profile in elliptical galaxies are isotropic at the centre and gradually becomes radial at the outskirts \citep{Cappellari07}. Our inferred anisotropy scaling factor $a_{\rm ani}$ is consistent within 1$\sigma$ with those reported by \citet{Birrer20} for both a sample \editrefth{of} time-delay lens galaxies and for a joint sample of time-delay lens galaxies and a subset of SLACS galaxies with integral-field unit spectroscopy (cf. table 6 in \citealt{Birrer20}). Note, there is significant overlap between our sample and the sample of \citet{Birrer20}, but the adopted mass models for the kinematics analysis are different. The associated uncertainties reported by \citet{Birrer20} are generally larger than our inference, which can be attributed to the mass model adopted by \citet{Birrer20}, that maximally explores the mass-sheet degeneracy.
}


\editfr{
In the next subsection, we place these two results in the context of elliptical galaxy formation and evolution.
}


\subsection{Evolution of massive elliptical galaxies} \label{sec:evolution}

\editfr{The current paradigm for the formation of elliptical galaxies consists of two phases \citep[e.g.,][]{Naab07, Guo08, Bezanson09, Furlong15}. In the first phase up to $z \approx 2$, gas condensates in a massive halo to form stars, and the dark matter distribution contracts as a result. In the second phase after $z\approx2$, elliptical galaxies grow in size \editfv{primarily} through multiple major or minor mergers. \editfv{Most of this later evolution is ``dry'' involving little gas and star formation, as evidenced by the almost uniformly old stellar populations in elliptical galaxies.} If elliptical galaxy halos contract up to $z\approx2$, our result then raises the question: how does the contracted halo at $z \approx 2$ expands back to resemble the NFW profile at $z\sim0.2$? Simulations have shown two principle mechanisms that can remove baryons from the central region and expand the dark matter halo -- AGN-driven gas outflow, and dynamical heating through accretion of materials from the environment \citep[e.g.,][]{Laporte12, Martizzi12}. Although mergers play a crucial rule in the growth and evolution of elliptical galaxies, simulations find that dissipationless mergers do not change the shape of the dark matter profile \citep{Gnedin04, Ma04}. Dissipational gas-rich mergers, in contrast, can make the profile steeper by transferring baryons towards the centre and thus further contracting the dark matter halo \citep{Sonnenfeld14}. The close-to-isothermal nature of the total mass distribution can also be explained as the end result of rearranging the mass distributions through accretion of collision-less materials in gas-poor mergers \citep{Johansson09b, Remus13}.
}

\begin{figure*}
	\includegraphics[width=\textwidth]{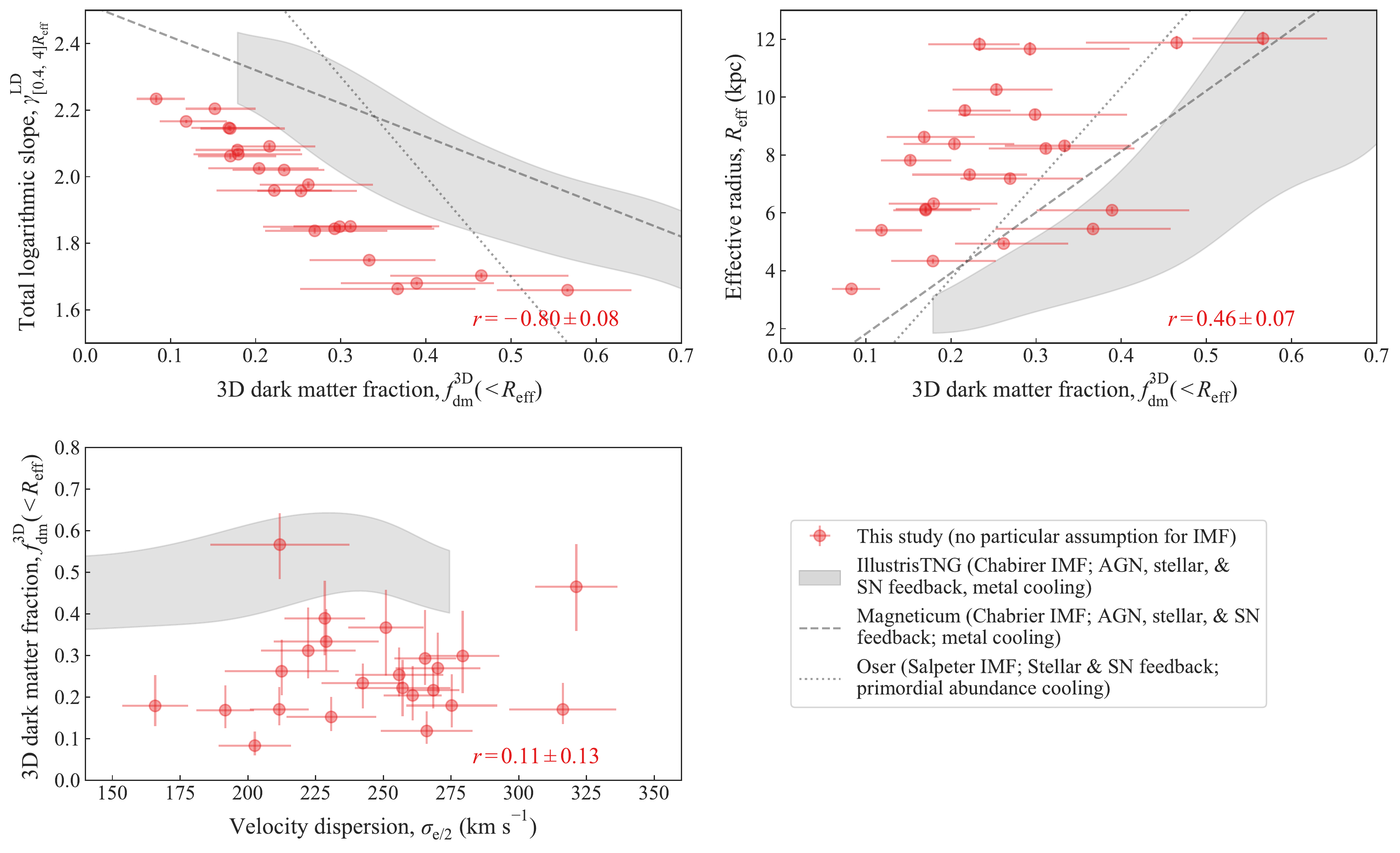}	
	\caption{ \label{fig:sim_compare}
	\editfr{Comparison of structural parameters of our SLACS sample with cosmological hydrodynamical simulations: IllustrisTNG \citep[grey shaded region \editfv{representing 1$\sigma$ confidence interval};][]{Wang20}, Magneticum \citep[grey dashed line,][]{Remus17}, and from \citet[][grey dashed-dotted line]{Oser10}. \editfv{The 1$\sigma$ interval for the IllustrisTNG simulation is produced by smoothing the corresponding distributions with a Gaussian kernel of bandwidth 0.05 for $f_{\rm dm}^{\rm 3D}$ and 20 km s$^{-1}$ for $\sigma_{\rm e/2}$.} \textbf{Top left:} distributions of the logarithmic slope $\gamma^{\rm LD}$ of 3D total mass profile within 0.4$R_{\rm eff}$ and 4$R_{\rm eff}$ and the 3d dark matter fraction $f^{\rm 3D}_{\rm dm}$ within $R_{\rm eff}$. The distributions from simulations do not match both the slope and the dark matter fraction simultaneously. \editrefth{The IllustrisTNG and Magneticum simulations adopted the Chabrier IMF, whereas our galaxies are consistent with the Salpeter IMF (Section \ref{sec:ml_gradient_discuss}). Adopting the Salpeter IMF would decrease $f_{\rm dm}$ in these simulation alleviating the disagreement.}} \editrefth{If the IMF mismatch, however, cannot explain all the observed differences (e.g., the Oser simulation adopted the Salpeter IMF and is still discrepant from our observed distribution), it would point to potential inadequacies in the implementation of baryonic physics in the simulation.} The orientation of our observed distribution is between the orientations from Magneticum and Oser simulations, which indicates moderate amount AGN and stellar feedback in SLACS galaxies' formation history. \textbf{Top right}: distributions of the 3D dark matter fraction $f^{\rm 3D}_{\rm dm}$ and the effective radius $R_{\rm eff}$. \editfv{These two quantities are moderately correlated, which is consistent with the scenario where elliptical galaxies grow in size through minor mergers and thus the central dark matter fraction within the half-mass radius increases.} \textbf{Bottom left:} distributions of the velocity dispersion within $R_{\rm eff}/2$ and $f_{\rm dm}^{\rm 3D}$. \editfv{The apertures for the observed velocity dispersions are adjusted following \citet{Jorgensen95}.} \editfv{The IllustrinsTNG galaxies match neither with the dark matter fraction distribution nor with the velocity dispersion distribution of their observational counterparts at the same stellar mass range.} 
	}
\end{figure*}

\begin{figure}
	\includegraphics[width=\columnwidth]{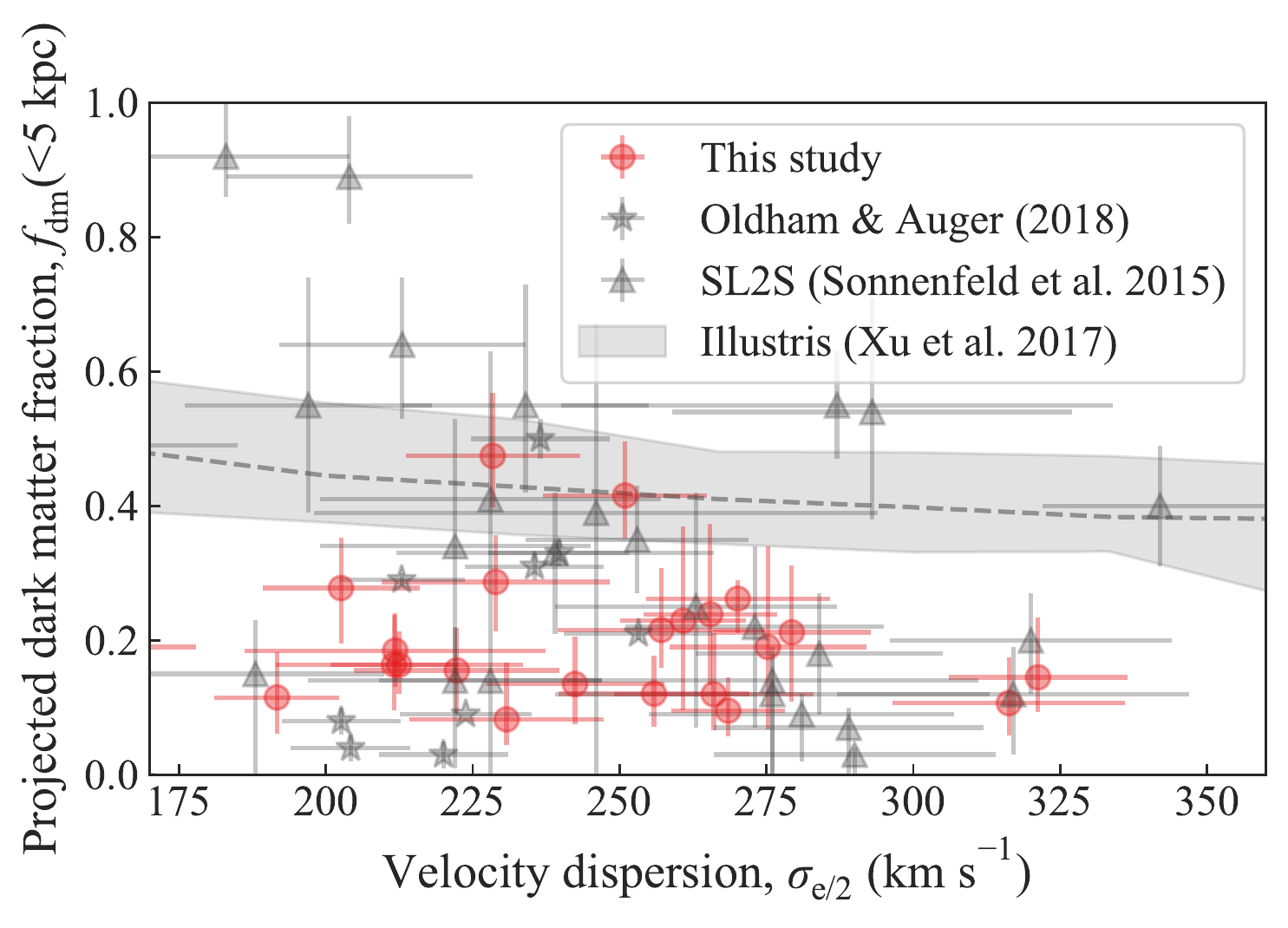}
	\caption{\label{fig:f_dm_lensing_compare}
	\editfr{Comparison of our observed distribution of projected dark matter fraction and the stellar velocity dispersion with previous observation and the Illustris simulation \citep{Xu17}. The grey shaded region shows the 90 per cent confidence interval and the dashed grey line shows the median of the Illustris distribution. Our observed distribution is qualitatively consistent with those from \citet{Sonnenfeld15} and \citet{Oldham18}. However, all of these strong-lensing based observations have lower average central dark matter fraction than predicted by the Illustris simulation. \editref{Note, the Illustris simulation adopts the Chabrier IMF. A Salpeter IMF would decrease $f_{\rm dm}$ in the Illustris simulation to bring the distribution closer to agreement with the observations.}}
	}
\end{figure}

\editfr{To investigate the driving mechanism behind the halo expansion observed in our sample, we compare the structural properties of our galaxies with those from simulations that adopt varying baryonic physics (Figure \ref{fig:sim_compare}). The simulation from \citet{Oser10} includes stellar and supernova feedback, and primordial abundance cooling. In contrast, the Magneticum and IllustrisTNG simulations includes AGN feedback in addition to stellar and supernova feedback, and the cooling mechanism includes metals \citep{Remus17, Weinberger17, Pillepich18}. For fair comparison with the simulations, \editfv{we adopt their definition and} compute the average logarithmic slope by fitting a power-law to the 3D density profile between 0.4$R_{\rm eff}$ and 4$R_{\rm eff}$. We also compute the central dark matter fraction within a 3D aperture of radius $R_{\rm eff}$. Here, we treat the half-light radius $R_{\rm eff}$ as the half-mass radius $R_{1/2}$, as the $M/L$ gradient exponent in our sample is very small and consistent with zero. The compared IllustrisTNG simulation corresponds to $z=0.2$, which is the mean redshift of the galaxies in our sample. The fitting functions from \citet{Remus17} corresponding to Magneticum and Oser simulations -- that are plotted in Figure \ref{fig:sim_compare} -- are shown to match with the simulated distributions over the redshift range $z=0$--2. In all of the panels of Figure \ref{fig:sim_compare}, the general trend found in the simulations \editfv{matches that in our observations}. However, the distributions themselves do not overlap. The mismatch \editfv{between our observation} and IllustrisTNG simulation predominantly arises from the higher $f_{\rm dm}^{\rm 3D}$ found in that simulation. However, the $\gamma$ distribution of the IllustrisTNG simulation is generally consistent with our sample. Similarly, the $\gamma^{\rm lensing}$ distribution obtained from fitting imaging-only data of simulated lenses in the SEAGLE simulation also reproduces the distribution from our observation (Figure \ref{fig:slope_dist}). However, the $f_{\rm dm}$ distribution of the SEAGLE simulation is not reported yet. In contrast, the Magneticum simulation produces higher radially averaged $\gamma$ distribution than our observed one, but reproduces a $f_{\rm dm}$ distribution that is generally consistent with our observation (cf. figure 8 of \citealt{Remus17}). \editrefth{Note, the simulations adopted different IMFs -- Chabrier IMF in the IllustrisTNG, EAGLE, and Magneticum simulations, and Salpeter IMF in the Oser simulation. Adopting the Salpeter IMF -- which is consistent with the galaxies in our sample (Section \ref{sec:ml_gradient_discuss}) -- would lead to lower $f_{\rm dm}$ in the corresponding simulations, thus bringing the $f_{\rm dm}$ closer to agreement with our observation. However, if the IMF mismatch cannot fully explain the observed differences, changes in the baryonic prescriptions in the simulations need to be fine tuned.} The slope of the correlation in our observed \editth{$f_{\rm dm}$--$\gamma$} distribution falls between the ones from the Magneticum and Oser simulations, which \editfv{is consistent with the fact} that the AGN feedback is a necessary driver in the evolution of these galaxies. To match the slope of the Magneticum simulation with our observation, weaker AGN and stellar feedback would be necessary. However, that would also push up the simulated $\gamma$ distribution and increase the discrepancy with our observation. Note that the lower values of $f_{\rm dm}$ than simulations is not unique to our lensing-based analysis, as previous lensing-based studies also find similarly lower $f_{\rm dm}$ distributions \citep[Figure \ref{fig:f_dm_lensing_compare};][]{Sonnenfeld15, Oldham18}. Moreover, the Magneticum and IllustrisTNG simulations themselves have offsets between their predicted $f_{\rm dm}$ distributions due to differences in their implementations of baryonic physics despite the similarities of the adopted feedback mechanisms.} 
\editfv{The offset between simulations highlights that the effect of AGN feedback depends critically on the specific implementation. Thus our observations cannot test AGN feedback in general, but they are a powerful empirical test of specific implementations.}

\editfr{
Similarly, the trend in our observed distribution of $f_{\rm dm}^{\rm 3D}$--$R_{\rm eff}$ matches with the ones from the simulations, however the distributions themselves do not match (Figure \ref{fig:sim_compare}). The moderate correlation ($r=0.46\pm0.07$) between the size of the galaxies and the central dark matter fraction supports that these galaxies have grown predominantly through minor mergers. As minor dissipataion-less mergers mostly deposit material at the outer region of the halo, the galaxy's half-mass radius increases as a result. Thus, the dark matter fraction within the half-mass radius also increases and becomes correlated with the half-mass radius. 
}

\editfr{The velocity dispersion distribution of the IllustrisTNG simulation is smaller than the measured distribution for our sample. \citet{Wang20} note that this difference between observation and simulation in velocity dispersion, along with the mismatch in the $f_{\rm dm}$ distribution, point to potentially insufficient implementation of baryonic physics in the IllustrisTNG simulation.
}

\editfr{
  In summary, our constraints on the elliptical galaxy structure in the context of previous observations and simulations \editfv{are consistent with} the following formation and evolution scenario for massive elliptical galaxies. \ttedit{The first stage of the formation of elliptical galaxies is through dissipational processes at $z\gtrsim 2$, when most of their present day stars are formed. The dissipation leads to contraction in the dark matter halo and the resultant total density profile observed in cosmological numerical simulations is steeper than the isothermal case \citep[e.g.,][]{Gnedin04, Naab07, Duffy10}. After $z \approx 2$, the growth of the elliptical galaxies is dominated by gas-poor dissipationless mergers, explaining their growth in size without the addition of younger stellar populations \citep[e.g.,][]{Newman12, Nipoti12}. This growth mechanisms is consistent with the $R_{\rm eff}$--$f_{\rm dm}$ correlation observed in our sample. Multiple dissipationless mergers decrease the total density profile of the galaxies to bring it close to isothermal, and increase the half-mass radius and the central dark matter fraction with decreasing redshift \citep{Tortora14}. Furthermore, AGN feedback expands back the contracted dark matter halos \citep{Martizzi13, Peirani19}, which is supported by the slope of our observed $\gamma^{\rm LD}_{[0.4,\ 4]R_{\rm eff}}$--$f_{\rm dm}^{\rm 3D}$ distribution. Dynamical heating from accretion may also play a role in expanding the dark matter halos in addition to the AGN feedback, however we do not find any indication either in favor or against the presence of dynamical heating in our sample. Gas-rich dissipational mergers can also happen at some point of the galaxies' growth history. However, the small intrinsic scatter in the halo response parameter ($\sigma_{\nu} \lesssim 0.1$) around the mean $\mu_{\nu} \sim 0$ in our sample, and the ages of the stellar populations \citep{Thomas05}, indicate that such gas-rich mergers are relatively rare, even though their contribution cannot be ruled out or confirmed conclusively given the present precision of numerical simulations and observations \citep[see, e.g.,][for discussion]{Sonnenfeld14,Remus17,Xu17}.}}




\subsection{Stellar IMF and $M/L$ gradient} \label{sec:ml_gradient_discuss}

\begin{figure}
	\includegraphics[width=\columnwidth]{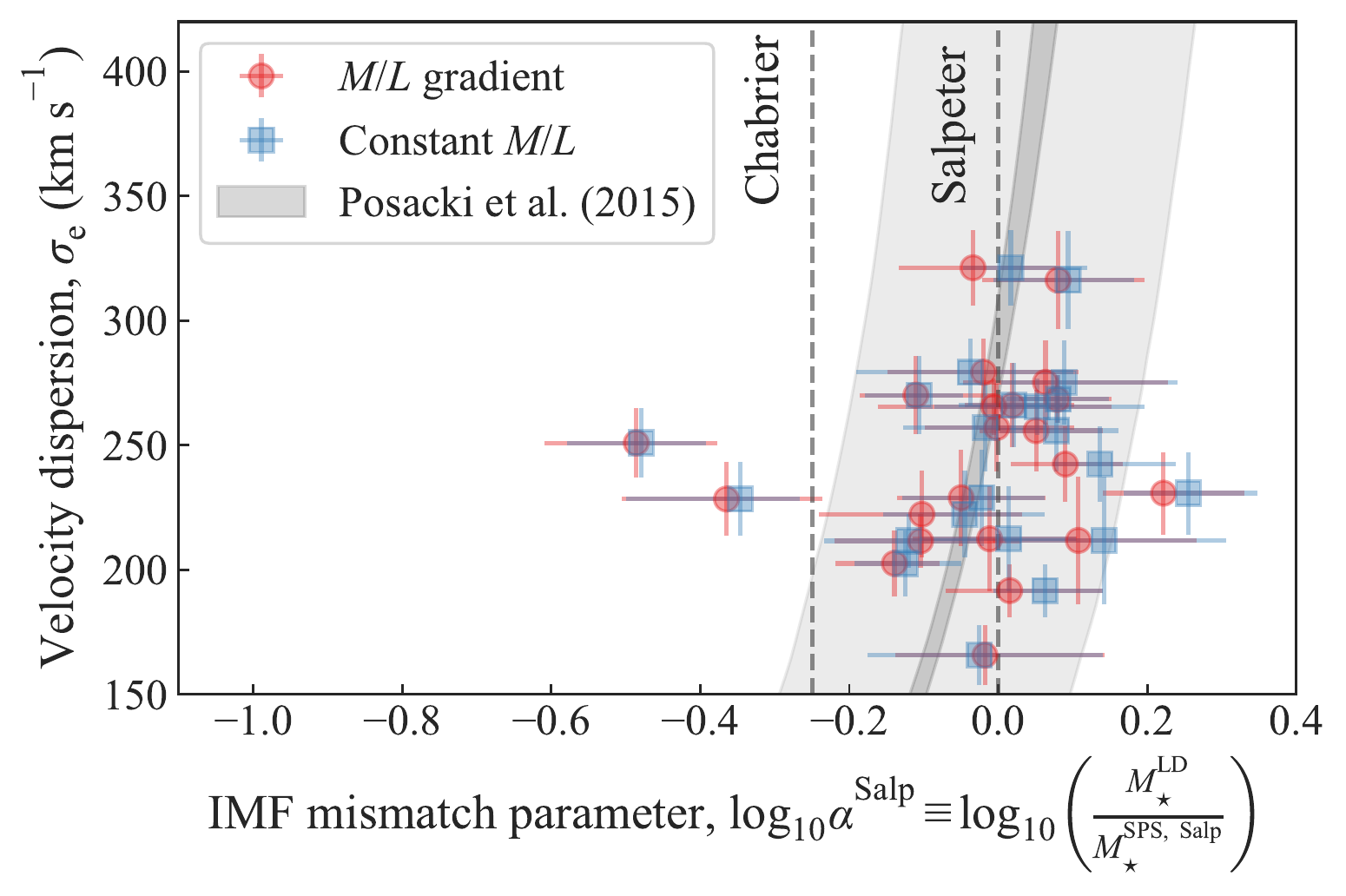}
	\caption{\label{fig:imf_compare}
	\editfr{Distribution of the IMF mismatch parameter $\alpha^{\rm Salp} \equiv M_{\star}^{\rm LD}/M_{\star}^{\rm SPS,\ Salp}$ and the velocity dispersion $\sigma_{\rm e}$ within the Effective radius $R_{\rm eff}$. The vertical dashed grey lines mark the $\alpha^{\rm Salp}$ values expected from the Salpeter or Chabrier IMFs. The distribution of IMF mismatch parameter is consistent with Salpeter IMF with median $\langle \log_{10} \alpha^{\rm Salp} \rangle_{\rm median} = 0.00 \pm 0.03$ and intrinsic scatter $0.11^{+0.03}_{-0.02}$. We also show the fitting function from \citet{Posacki15} obtained from fitting SLACS \citep{Auger10b} and ATLAS$^{\rm 3d}$ galaxies. The deeper grey shaded region represent the mean of the relation and the lighter grey shaded region represents the rms scatter around the relation.}
	}
\end{figure}

\editfr{
Our inferred stellar mass $M_{\star}^{\rm LD}$ distribution from the joint lensing--dynamics analysis support a heavier stellar IMF such as the Salpeter IMF. In Figure \ref{fig:imf_compare}, we illustrate the distribution of the IMF mismatch parameter as defined by \citet{Treu10} as
\begin{equation}
	\alpha^{\rm Salp} \equiv \frac{M_{\star}^{\rm LD}}{M_{\star}^{\rm SPS,\ Salp}},
\end{equation}
where $M_{\star}^{\rm SPS,\ Salp}$ is the SPS-based measurement of the stellar mass assuming the Salpeter IMF. If the true underlying IMF follows the Salpeter function, then $\alpha^{\rm Salp} = 1$ is expected. For the Chabrier IMF, in comparison, the IMF mismatch parameter should be $\alpha^{\rm Salp} \approx 0.56$ or $\log_{10} \alpha^{\rm Salp} \approx -0.25$. The mean IMF mismatch parameter of our sample is $\langle  \log_{10} \alpha^{\rm Salp} \rangle_{\rm median} = 0.00 \pm 0.03$ with an intrinsic scatter of $0.11^{+0.03}_{-0.02}$. Our observed distribution of $\alpha^{\rm Salp}$--$\sigma_{\rm e}$ is in good agreement with the fitting function provided by \citet{Posacki15} 
\begin{equation}
	\log_{10} \alpha^{\rm Salp} = (0.38 \pm 0.04) \times \log_{10} \left( \frac{\sigma_{\rm e}}{200\ {\rm km\ s^{-1}}}\right) + (-0.06 \pm 0.01 ),
\end{equation}
which is obtained from fitting SLACS \citep{Auger10b} and ATLAS$^{\rm 3D}$ \citep{Cappellari13} galaxies. This relation has a root-mean-square (rms) scatter of 0.12 dex.
}

\editfr{Several previous strong-lensing studies also found evidence for heavier IMF such as the Salpeter} \citep[e.g.,][]{Treu10, Spiniello11, Spiniello12, Sonnenfeld12, Oldham18}. \editth{Multiple non-lensing studies similarly found evidence for a heavy IMF in elliptical galaxies -- e.g., from dynamics of local elliptical galaxies \citep{Cappellari12}, and based on single stellar population (SSP) models \citep[e.g,][]{Conroy12, LaBarbera13, Spiniello14}.}

\editfr{
Whereas the observed IMF in the Milky Way \editfv{stellar populations} is closer to a Chabrier IMF \citep[e.g.,][]{Bastian10}, our observation of a Salpeter-like IMF in massive elliptical galaxies indicates that stars formed in a different environment in these elliptical galaxies. Star formation in a turbulent environment with high gas density -- that can result from gar-rich mergers and accretion events at the early phase of elliptical galaxy formation -- can lead to a heavy IMF such as the Salpeter IMF \citep{Hopkins13, Chabrier14}.
}
 

\editfr{We find that the $M/L$-gradient exponent is very small and consistent with zero, as the 95 per cent upper limit for the sample mean of the $M/L$-gradient exponent $\eta$ is 0.02.} \editth{This small value is consistent with \citet{Oldham18}, who found that most of their lens galaxies individually favor $\eta < 0.1$. \editfr{Our result, however, is in tension with \citet{Sonnenfeld18b}, who inferred $\mu_{\eta}=0.24 \pm 0.04$ and $\sigma_{\eta}=0.09\pm0.02$ from a larger SLACS subsample and HSC weak-lensing information. Some fraction of the difference can be explained by the difference in the adopted light profile. Whereas \citet{Sonnenfeld18b} adopt a de Vaucouleurs profile for $I$ band, we adopt a double S\'ersic profile in V-band. The double S\'ersic profile fits the light distribution better than the de Vaucouleurs profile. We observe that our double S\'ersic profile fits are generally steeper than the de Vaucouleurs fit of the SLACS galaxies from \citet{Auger09}, which explains the steeper mass profile than the light profile reported by \citet{Sonnenfeld18b}. Such steeper mass profile than the light profile can not arise from a color gradient between $I$ and $V$ bands, as observationally $I$-band light profile is steeper than the $V$ band as the color gets bluer radially outward \citep[e.g.,][]{Tamura00}. Furthermore, \citet{Sonnenfeld18b} adopt an isotropic profile for orbital anisotropy in their dynamical model, whereas we adopt an Osipkov--Merritt anisotropy profile, which can potentially account for some part of the observed discrepancy \editfv{by trading radial anisotropy for $M/L$ gradient to reproduce the observed velocity dispersion}.}
}

\editfr{
Several recent studies based on SPS modelling have pointed to a radially varying IMF in local elliptical galaxies, where the IMF in the central $\lesssim2$ kpc region is bottom-heavy -- even super-Salpeter -- and it gets more bottom-light radially outward to match with the Chabrier IMF in the outer region \citep[e.g.,][]{MartinNavarro15, vanDokkum17, LaBarbera19}. Such a radially varying IMF is consistent with the scenario where the central stellar population is dominated by \textit{in situ} stars that formed following the Salpeter IMF in a highly turbulent environment at high redshift ($z \gtrsim 2$), and the stellar population in the outer region mostly comes from merged satellites that had a different star-forming environment closer to the Milky Way and thus the Chabrier IMF. However, this radially varying IMF is interpreted from a radially varying $M/L$ ratio based on the SPS models. Thus, at face value our result of very small $M/L$ gradient appears to be in tension with the SPS-based studies mentioned above. However, the $M/L$ gradient observed in the SPS-based studies is dominated by a prominent gradient at $<0.1R_{\rm eff}$, whereas our joint lensing--dynamics analysis is sensitive near $0.5R_{\rm eff}$--$R_{\rm eff}$ where the SPS-based $M/L$ flattens out. Thus, the apparent difference between our result and SPS-based $M/L$ gradient results can be reconciled. Indeed, \citet{vanDokkum17} point out that the observed Salpeter IMF in joint lensing--dynamics studies is consistent with the SPS-based result of a radially varying IMF, if the measurement scales are taken into account [cf. Figure 17 of \citet{vanDokkum17}].
}

\editref{Our inferred star formation efficiency is larger by approximately a factor of 2 than those reported by studies using abundance matching at the relevant redshift and halo mass assuming the Chabrier IMF \citep{Behroozi13, RodriguezPuebla17, Girelli20}. Adopting the Salpeter IMF would make the star formation efficiency derived from abundance matching consistent with our result.}

\subsection{Implication for time-delay cosmography} \label{sec:time_delay_cosmography}

\editfr{In time-delay cosmography, the measurement of the Hubble constant $H_0$ requires an accurate constraint \editfv{on the surface density profile} near the Einstein radius. Similar to this study, analyses by the Time-Delay COSMOgraphy (TDCOSMO) have adopted a power-law mass distribution -- among other choices -- for lens modelling \citep{Suyu10, Suyu13, Wong17, Birrer19, Chen19, Rusu20, Shajib20}. A sample of seven strong-lensing systems with measured time delays provide a $\sim 2$ per cent measurement of the Hubble constant \citep{Millon20}. However, relaxing the assumption on the power-law profile and thus allowing the full range of mass-sheet degeneracy in the lens models inflates the uncertainty to \editfv{$\sim$8} per cent \citep{Birrer20}. As a result, precise knowledge of the internal structure of elliptical galaxies is currently the limiting factor for the time-delay measurement of $H_0$. Therefore, one option to shrink back the uncertainty on $H_0$ is \editsv{to apply} an external prior on the internal structure of time-delay lens galaxies, and thus \editsv{narrow} down the allowed range of the \editsv{MSD}. \editfv{Alternatively,} \citet{Birrer20} constrain the MSD by incorporating the structural information from the power-law lens models presented in this study and their measured kinematics \editsv{{\it under the assumption that the TDCOSMO and SLACS lenses are drawn from the same parent population.} The additional hypothesis and the addition of external information reduces the uncertainty on $H_0$ to 5 per cent.} However, including the structural information from the SLACS galaxies also increases the mean surface density of the time-delay lens galaxies near the Einstein radius by \editfv{$\sim$9.5} per cent from the power-law estimate, which also translates to a \editfv{$\sim9.5$} per cent decrease in the inferred $H_0$. This increased estimate of the surface density is consistent \editsv{within the uncertainty} with our joint lensing--dynamics analysis, as we find on average $\sim$5 per cent higher surface density near the Einstein radius for ours stars+NFW model than the power-law model (see Figure \ref{fig:deviation_from_power_law}).
}

\editfr{\editsv{TDCOSMO} analyses have also adopted an NFW+stars model in addition to the power-law model and found that the two models are in very good agreement \citep{Millon20}. In contrast, we observe that the NFW+stars model deviate from the power law on average by $\sim$5 per cent \editsv{at the Einstein radius}. We show that the deviation of the NFW+stars model from the power-law model in our analysis is largely driven by the adopted prior on the halo mass. As strong-lensing and kinematics data mostly constrain the mass distribution in the central region, the total halo mass or the dark matter profile is poorly constrained from lensing--dynamics analysis without a prior on the halo mass, or on the mass distribution at scales larger than the NFW scale radius. Our empirical prior on the halo mass $M_{200}$ is obtained from the HSC weak-lensing measurements of the SDSS galaxies, which is the parent sample of the SLACS galaxies. Moreover, the selection function for the SLACS galaxies is accounted for in this prior. If instead, we \editsv{had} adopted a prior on $M_{200}$ \editsv{with} a smaller mean by $\sim$0.3 dex, the resultant NFW+stars density profile would \editsv{have matched} the power-law model very well on average (Figure \ref{fig:deviation_from_power_law}). A higher $M_{200}$ translates to a higher normalization of the NFW profile, thus the total density profile becomes dark-matter-dominated at a relatively smaller radius and the slope of the total density profile gets shallower. We provide two possible explanations for why our NFW+stars model deviates from the power-law whereas the ones in the \editsv{TDCOSMO} analyses do not.
}

\editfr{The first possible explanation is that there may be a bias in the NFW profile normalization in the \editsv{TDCOSMO} models due to \editsv{weak prior constraints on} the mass distribution at large scales. \editsv{TDCOSMO} analyses have generally adopted a prior only on the NFW scale radius, which only has a weak constraint on $M_{200}$ as the $M$--$c$ relation is not imposed. As a result, the resultant halo mass may be biased \editsv{low} as lens models tend to produce lower halo masses without strong priors on the mass distribution at scales larger than the NFW scale radius \citep[e.g.,][]{Oldham18}. A smaller halo mass by $\sim$0.3 dex may not be noticeable within the measurement uncertainty when looking at individual lens galaxies, \editsv{but it} may be identifiable by investigating the $M_{\star}$--$M_{\rm 200}$ relation for the \editsv{TDCOSMO} lenses. \editfv{However}, due to the typically large intrinsic scatter in $M_{200}$, a sample size of seven may not be sufficient to identify such a bias.
}

\editfr{The second possible explanation is that the SLACS lens galaxies \editfv{and the \editsv{TDCOSMO} lens galaxies do not belong to the same galaxy population. Even if the predecessors of the SLACS galaxies at $z=0.6$ represent the time-delay lens galaxy population, they may have evolved since $z=0.6$ to have a different internal structure at $z=0.2$ \editsv{\citep[see, e.g., ][]{Sonnenfeld15}}.} A $M_{200}$ prior with larger mean leads to a larger dark matter fraction $f_{\rm dm}$. As $f_{\rm dm}$ is observed to increase from higher redshift to lower redshift \citep{Tortora14}, it is possible that a lower $f_{\rm dm}$ in the time-delay lenses with mean deflector redshift $\langle z_{\rm d}\rangle=0.6$ makes the power-law model and the NFW+stars model to be in good agreement.} 

\editfr{We note that an increasing $f_{\rm dm}$ with decreasing redshift in the \editsv{TDCOSMO} lenses would also be consistent \editsv{at least qualitatively} with the weak trend observed in the individual $H_0$ measurements from these lenses \citep{Wong20}. As increasing $f_{\rm dm}$ can create larger positive deviation between the NFW+stars model and the power-law model, the resultant $H_0$ from the power-law model would shift towards higher values with decreasing redshift. \editfv{However}, this scenario requires a combination of both of the above explanations where high-redshift galaxies are less vulnerable to an NFW normalization bias due to their lower underlying $f_{\rm dm}$, and lower redshift galaxies are more vulnerable to the normalization bias due to their higher underlying $f_{\rm dm}$.
}

\editfr{
\citet{Blum20} suggest that a hypothetical 10 per cent positive bias in the Hubble constant from time-delay cosmography would point to the presence of a large core in the dark matter distribution. We show that such a shift in the Hubble constant can also be explained by a relatively higher normalization in the vanilla NFW profile, if the SLACS lens galaxies and \editsv{the TDCOSMO} lens galaxies are structurally self-similar.
}




\subsection{Limitations of this study} \label{sec:limitations}

\edits{Although we do not account for external convergence from the line-of-sight structures \editfr{in our \editfv{lens models}}, \ttedit{this simplification has no impact on our results, as we discuss in this paragraph.} \citet{Birrer20}} \editfv{find the mean external convergence for a subsample of 33 SLACS lenses to be $\langle \kappa_{\rm ext}\rangle = 0.005_{-0.009}^{+0.022}$. However, this subsample is curated to only select systems with small local overdensities [see \citet{Birrer20} for details]. Our sample of 23 lenses has a large overlap with this sample of 33 lenses from \citet{Birrer20}, thus the overlapping fraction of our sample has negligible mean external convergence. The external convergences for the remaining systems are likely to be not negligible and our joint lensing--dynamics analysis can be suspect of bias due to not correcting for the external convergence in our joint lensing--dynamics analysis. However, we only adopt MSD-invariant quantities that are not sensitive to the external convergence -- such as the Einstein radius $R_{\rm E}$ and the local differential term $R_{\rm E} \alpha^{\prime\prime}_{\rm E}/(1 - \kappa_{\rm E})$ -- as the lensing constraints in our joint lensing--dynamics analysis. Thus, our joint lensing--dynamics analysis is insensitive to the external convergence.}


\editfr{
In our adiabatic contraction model, we assume that the stellar mass distribution initially resembled the NFW profile following the adiabatic contraction model of \citet{Blumenthal86}. This is certainly not true as the stellar mass delivered to the elliptical galaxies through mergers did not resemble the NFW profile at the time of the merger. However, cosmological hydrodynamical simulations have shown that such a simple model can reproduce the contracted profile of the dark matter even after gas-rich mergers, albeit with a modification in the adiabatic invariant quantity $rM(r)$ \citep[e.g.,][]{Gnedin04, Duffy10}. We adopted the modification of \citet{Dutton07}, which can approximate the contracted profiles of both \citet{Gnedin04} and \citet{Abadi10}. However, since cosmological hydrodynamical simulations have so far been unable to match all the properties of their observational counterparts, our simple model for adiabatic contraction may not be fully justified to truthfully represent the interplay between the baryonic and the dark matter distributions. Such inconsistencies may be identified by an alternative analysis of the same sample that adopts the generalized NFW profile for the dark matter distribution, and compare the results with the ones from our the adiabatic contraction model. We leave such explorations for future studies.
}

\section{Summary} \label{sec:summary}

We uniformly modelled a sample of \numlens\ SLACS lenses and constrained their structural properties from a joint lensing--dynamics analysis. The lens modelling in this study is different from the original SLACS analysis, in which the lens images were modelled by fixing the logarithmic slope to $\gamma=2$ and then the logarithmic slopes were inferred from the stellar kinematics. In contrast, in this study we first estimate the logarithmic slopes only from the lensing observables, i.e., the lens image. We then combine the stellar kinematics to constrain the amount of contraction in the dark matter distribution for two models of stellar mass distribution: (i) with constant $M/L$ and (ii) with $M/L$ gradient. We summarize the main results of this paper below.
\begin{itemize}
	\item \editf{From the combination of lensing and kinematic observables, \editth{we constrain the average halo response parameter \editet{$\mu_{\nu} = -0.06 \pm 0.04$ with intrinsic scatter $\sigma_{\nu} \leq 0.092$ (95 per cent upper limit)} for a constant stellar $M/L$ model. For a stellar $M/L$ gradient model, we find \editet{$\mu_{\nu} = -0.03^{+0.04}_{-0.05}$ and $\sigma_{\nu} \leq 0.074$}.} \ttedit{Our results are consistent with a dark matter halo described by an NFW profile with no contraction \edits{nor expansion}.}} \editfr{For comparison, the \citet{Blumenthal86} model corresponds to $\nu = 1$, the contraction in \citet{Gnedin04} simulations correspond to $\nu = 0.8$, and the contraction in \citet{Abadi10} simulations correspond to $\nu \sim 0.4$ -- which are all ruled out by our result. \ttedit{Our results are consistent with a scenario in which elliptical galaxies grow \editet{by dissipational processes} at $z\gtrsim2$, steepening their dark matter halos. At later times, AGN feedback -- with potential additional contributions from dynamical heating through accretion events -- expand the dark matter halos back to an NFW profile, on average.}}
	\item The distribution of logarithmic slopes $\gamma$ \editfv{for the power-law model} constrained from the imaging\editfv{-only} data has a median $\langle \gamma^{\rm lensing}_{\rm PL} \rangle = 2.08\pm0.03$ and an intrinsic scatter $0.13 \pm 0.02$. \editf{This is consistent with the slope distribution from the SLACS analysis $\langle \gamma^{\rm LD}_{\rm PL} \rangle = 2.078 \pm 0.027$ with an intrinsic scatter of $0.16\pm0.02$.} \editfv{We find that the NFW+stars profile constrained from our joint lensing--dynamics analysis only deviates by $\lesssim 5$ per cent on average near the Einstein radius \editsv{($R_{\rm E} \sim R_{\rm eff}/2$ for our sample)}, with \editsv{even smaller deviation at smaller scales}. The small deviation in the mean compared to the intrinsic scatter explains the good agreement between the average local logarithmic slope from lensing-only data and the radially averaged logarithmic slopes from \citet{Auger09}.} 

	\item \editth{For the stellar $M/L$ gradient model, we find that most galaxies \ttedit{do not require a significant $M/L$ gradient}. The 95 per cent upper limit for the sample mean of the exponent $\eta$ \editsv{in our $M/L$-gradient model $\Upsilon \propto (R/R_{\rm eff})^{-\eta}$} is 0.02, \ttedit{\editsv{which corresponds to $\lesssim$5 per cent decrease in $M/L$ between $0.1R_{\rm eff}$ and $R_{\rm eff}$}}.} \editfr{Moreover, the inferred stellar masses from joint lensing--dynamics analysis for the galaxies in our sample is consistent with the Salpeter IMF, with the IMF mismatch parameter $\langle \log_{10} \alpha^{\rm Salp}\rangle_{\rm median} = 0.00 \pm 0.03$ with an intrinsic scatter $0.11^{+0.03}_{-0.02}$. Such a heavy IMF in the central regions of the elliptical galaxies can be explained by a different star-forming environment than in the Milky Way, e.g., the presence of turbulence in high gas density.}
\end{itemize}

\editfv{In the future, larger samples of galaxy--galaxy lenses at different redshifts will be able to further constrain the evolutionary tracks of elliptical galaxies. Such larger samples can be assembled from past surveys with available high-resolution imaging and ancillary data, e.g., the full SLACS sample \citep{Auger09}, the Strong Lensing Legacy Survey (SL2S) sample \citep[][]{Sonnenfeld13}, and the SLACS for the MASSES (S4TM) sample \citep{Shu17}. Current surveys such as the Dark Energy Survey (DES) and the Dark Energy Spectroscopic Instrument (DESI) Legacy Imaging Surveys are producing new galaxy--galaxy lens candidates for confirmation and follow-up on the order of hundreds \citep{Jacobs19, Jacobs19b, Huang20b}.}
\editf{Future surveys, e.g., the Vera Rubin Legacy Survey for Space and Time, the Nancy Grace Roman Space Telescope, \editref{and \textit{Euclid}}, will \editfv{increase the number of newly discovered galaxy--galaxy lenses to thousands} \citep{Collett15}. 
} \editfv{An automated and uniform modelling pipeline will be essential to study such large samples of lenses and this paper has taken the initial steps towards such an automated pipeline.}

\section*{Acknowledgements}
We thank the anonymous referee for many useful comments that helped us improve this manuscript. We thank Matthew Auger for sharing the SLACS weak lensing \editfv{measurements from \citet{Auger10} in digital form}. We express gratitude to Adriano Agnello, Elizabeth Buckley-Geer, Thomas Collett, Frederic Courbin, Xuheng Ding, Aymeric Galan, Martin Millon, Veronica Motta, Sampath Mukherjee, Dominique Sluse, and Chiara Spiniello for providing suggestions that improved this analysis and manuscript. We additionally thank Matthew Auger, Chris Fassnacht, and Leon Koopmans for helpful discussions. \editf{We also thank the SLACS team for collecting the wonderful data used in this paper.} AJS and TT were supported by the National Aeronautics and Space Administration (NASA) through the Space Telescope Science Institute (STScI) grant HST-GO-15320. AJS was additionally supported by a Dissertation Year Fellowship from UCLA Graduate Division. This research was supported by the U.S. Department of Energy (DOE) Office of Science Distinguished Scientist Fellow Program. TT acknowledges support by the Packard Foundation through a Packard Research fellowship \editfr{and by the National Science Foundation through NSF grants AST-1714953 and AST-1906976}. The development of \textsc{dolphin} is supported by NASA through the STScI grant HST-AR-16149.

This work used computational and storage services associated with the Hoffman2 Shared Cluster provided by UCLA Institute for Digital Research and Education’s Research Technology Group. AJS thanks Smadar Naoz for providing access to additional computing nodes on the Hoffman2 Shared Cluster.

This research made use of \textsc{lenstronomy} {\citep{Birrer15, Birrer18}}, \textsc{dolphin} (\url{https://github.com/ajshajib/dolphin}), {\textsc{fastell} \citep{Barkana99}}, \textsc{numpy} \citep{Oliphant15}, \textsc{scipy} \citep{Jones01}, \textsc{astropy} \citep{AstropyCollaboration13, AstropyCollaboration18}, \textsc{jupyter} \citep{Kluyver16}, \textsc{matplotlib} \citep{Hunter07}, \textsc{seaborn} \citep{Waskom14}, \textsc{sextractor} \citep{Bertin96}, \textsc{emcee} \citep{Foreman-Mackey13}, \textsc{colossus} \citep{Diemer18}, \textsc{pomegranate} \citep{Schreiber18}, and \textsc{chainConsumer} (\url{https://github.com/Samreay/ChainConsumer}).

\section*{Data availability}

The \textit{HST} imaging data used in this paper are publicly available from Mikulski Archive for Space Telescopes (MAST). The other ancillary measurements -- i.e., the stellar kinematics and the weak lensing measurements -- are obtained from previous studies and we provide references to these previous studies. These measurements can be obtained either directly from the paper or by requesting the corresponding author of the related paper. The lens modelling codes \textsc{lenstronomy} and \textsc{dolphin} used in this paper are publicly available on GitHub.





\bibliographystyle{mnras}
\bibliography{../../ajshajib_old}



\appendix

\section{Mass and light alignments} \label{app:alignments}

Alignment between the dark and baryonic components of mass distribution can be used to validate the predictions from simulations. In elliptical lens galaxies, the misalignment between the mass and light is observed to be within $\pm$10$\degr$ in the absence of large external shear effects ($\gtrsim$0.1), whereas larger misalignments are usually accompanied with large external shear \citep[e.g.,][]{Keeton98, Koopmans06, Treu09, Sluse12, Shajib19}. This observation agrees well with the Illustris simulation \citep{Xu17}. \editfr{However, there have been mismatching reports in the literature on the correlation between the light and mass ellipticities or axis ratios, which can most likely be attributed to different selection functions. \editfr{The differences in the selection functions can arise from different lens-finding methods and from different population of lenses -- e.g., quads or doubles, galaxy--galaxy lenses or lensed quasars.} For example, \citet{Koopmans06, Sluse12, Gavazzi12} report strong correlation between mass and light ellipticities. In comparison, \citet{Keeton98, Ferreras08, Rusu16, Shajib19} find weak to no correlation between the mass and light ellipticities.} 

\editfr{In this appendix, we investigate the alignment between mass and ellipticities and their correlation with some other model parameters. We present the results in Appendix \ref{sec:alignment} and discuss them in Appendix \ref{sec:alignment_discussion}.}

\subsection{Results} \label{sec:alignment}

\begin{figure*}
	\includegraphics[width=\textwidth]{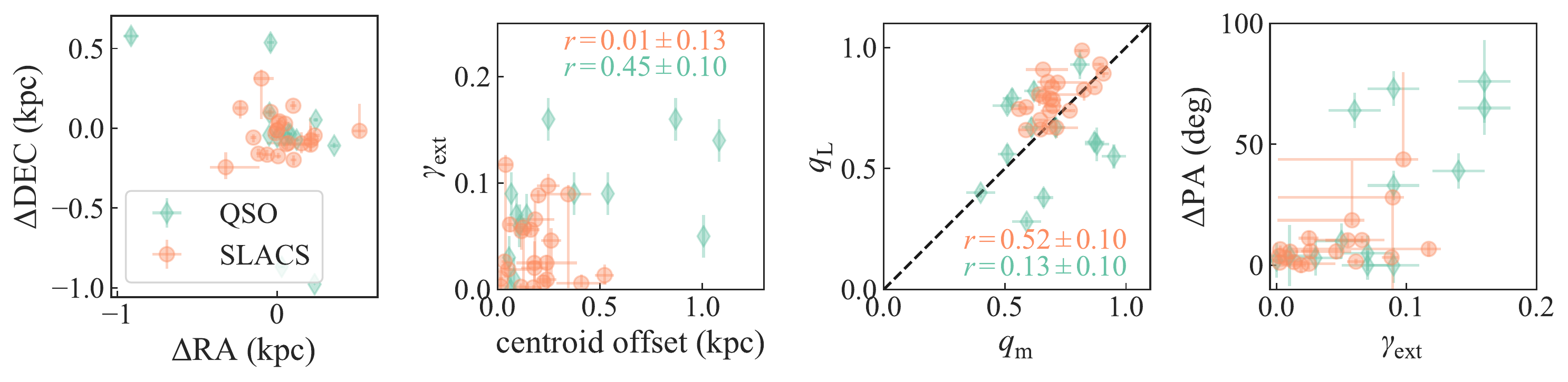}
	\caption{ \label{fig:mass_light_alignment}
	The alignment between mass and light distribution for SLACS lenses in our sample (orange) and quasar lenses from \citet[][green]{Shajib19}. \textbf{First panel:} The angular offset between the centroids of mass and light. For our fiducial cosmology, the offset distribution for the SLACS lenses has a 95 per cent upper limit of $426\pm68$ pc. \editth{\textbf{Second panel:} Distribution of the absolute centroid offsets and external shear. These two quantities are moderately correlated for the quasar lenses from \citet{Shajib19} with biweight mid-correlation $r=0.45 \pm 0.10$, which indicates that the presence of nearby perturbers may cause an offset between mass and light centroids in the quasar lenses. However, there is no such correlation for the SLACS lenses. Although the shear distribution for the SLACS lenses is at a lower range than that of the quasar lenses, thus any such potential correlation is hard to identify given the measurement uncertainties.} \textbf{Third panel:} The distribution of mass and light axis ratios. The axis ratios are weakly correlated with biweight mid-correlation \editref{$r=0.52\pm0.10$}. \textbf{Fourth panel:} The distribution of major axis misalignment between mass and light and external shear. \editsv{The 19 lenses from our sample of 23 that have $q_{\rm L} < 0.9$ and $q_{\rm m} < 0.9$ -- and thus reliable position angle estimates -- are plotted in this panel.} For these SLACS lenses, misalignment angle $\Delta$PA $> 12$ deg are the systems with $\gamma_{\rm ext} \gtrsim 0.05$.
	}
\end{figure*}

We illustrate the distributions of these quantities in Figure \ref{fig:mass_light_alignment}. \editfr{In the following subsections, we present the offset between the centroids and the misalignments between ellipticity magnitudes and orientations.}

\subsubsection{Centroid}

We find an rms scatter of $67 \pm 5$ \editf{mas} in the centroid offset. \editth{For our fiducial cosmology, the 68 per cent and 95 per cent upper limits of the absolute offsets are $218 \pm 19$ pc and $426 \pm 68$ pc, respectively (Figure~\ref{fig:mass_light_alignment}, first panel).} 

\subsubsection{Ellipticity}
We find a weak correlation between the axis ratios of the mass and light distributions with a biweight mid-correlation of \editref{$r=0.52\pm 0.10$} (Figure \ref{fig:mass_light_alignment}, \editref{third} panel). Biweight mid-correlation is similar to the Pearson's $r$ coefficient. However, it depends on the median instead of the mean, thus it is more robust against outliers \citep[e.g.,][]{Beers90}. We get the uncertainty in the mid-correlation by sampling the axis ratios of all the lenses 1000 times from their posterior distributions.

\editth{We check if the lens systems are double-image or quadruple-image systems by checking the number of images produced by the lens models for a hypothetical point source located at the source centres. We find 19 out of 23 lenses to be double-image systems and the remaining 4 to be quadruple-image systems.} \editfv{The majority of doubles explain the lower ellipticity and shear distributions observed in the SLACS sample compared to that of the quadruply lensed quasars (Figure \ref{fig:mass_light_alignment}).}

\subsubsection{Position angle}

While comparing the position angles of the mass and light, we ignore 4 lenses with $q_{\rm m} > 0.9$ and $q_{\rm L} > 0.9$, as the estimate of the position angle in low-ellipticity cases can be unreliable. Out of the remaining 19 lenses, we find 16 to have misalignment angle $\Delta$PA$\leq12$ deg. The three systems with $\Delta$PA$>12$ deg have relatively higher external shear $\gamma_{\rm ext} \gtrsim 0.05$ (Figure \ref{fig:mass_light_alignment}, right-hand panel).

\subsection{Discussion} \label{sec:alignment_discussion}
We find moderate correlation between axis ratios of the mass and light (biweight mid-correlation $r=0.55\pm0.09$). Although \citet{Gavazzi12} and \citet{Sluse12} both report strong correlation between mass and light ellipticities, we compute the biweight mid-correlation from the values reported by these authors to find $r=0.26 \pm 0.07$ (weak correlation) from the galaxy--galaxy lenses of \citet{Gavazzi12} and $r=0.91 \pm 0.02$ (very strong correlation) from the lensed quasars of \citet{Sluse12}. \editth{Additionally, \citet{KostrzewaRutkowska14} find strong correlation ($r = 0.77 \pm 0.16$) from a sample of 9 galaxy--galaxy lenses from the Cambridge And Sloan Survey Of Wide ARcs in the skY (CASSOWARY).} In contrast, \citet{Shajib18} find very weak to no correlation between mass and light for a sample of 13 quadruply lensed quasars ($r=0.13 \pm 0.10$). Similarly, \citet{Keeton98}, \citet{Ferreras08}, and \citet{Rusu16} also find no correlation between the mass and light ellipticities. \editth{Some of the differences between these studies, including ours, can be due to the differences in the data quality, analysis techniques, and adopted models.} \edits{Some of these differences can also be due to the selection function, {for example between the galaxy--galaxy lenses and lensed quasars}. SLACS is deflector-selected and mostly comprises doubly lensed objects, whereas quadruply lensed quasars are expected to be strongly selected to favor big inner caustics \editth{\citep{Dobler08, Collett16}}. Therefore, they favor high ellipticity and shear. Moreover, the SLACS sample has a smaller average redshift than the sample of quasar lenses, e.g., from \citet{Shajib19}. Thus, the line-of-sight effect is much less important for SLACS than for the lensed quasars.} \editfv{As \citet{Shajib19} model their quasar lens sample using \textit{HST} imaging of comparable quality, similar models, and the same modelling software as in this paper, the observed differences between the SLACS sample and the quasar lens sample in Figure \ref{fig:mass_light_alignment} have to arise from the differences in the selection functions.}

We find that in most galaxies with moderate ellipticity ($q \leq 0.9$), the major axes of the mass and light distributions are well aligned within $\pm 12$ deg. The lenses with large misalignment ($\Delta$PA$>12$ deg) also have relatively larger ($\gamma_{\rm ext} \gtrsim 0.05$) external shear within the sample. This result is consistent with previous studies  \citep{Kochanek02, Ferreras08, Treu09, Gavazzi12, Sluse12, Bruderer16, Shajib18}. The absence of systems with large misalignment angle and low external shear is consistent with the prediction of galaxy formation models that highly misaligned orbits in isolated galaxies are unstable and thus rare \citep[e.g.,][]{Adams07, Debattista15}. Only in blue starburst galaxies -- unlike the galaxies in our sample -- constant gas-flow can sustain highly misaligned orbits \citep{Debattista15}.

\section{Algorithm to mask the lensed arcs} \label{app:arc_mask}

\begin{figure}
	\begin{subfigure}[t]{\columnwidth}
		\begin{minipage}{.35\columnwidth}
			\includegraphics[width=\columnwidth]{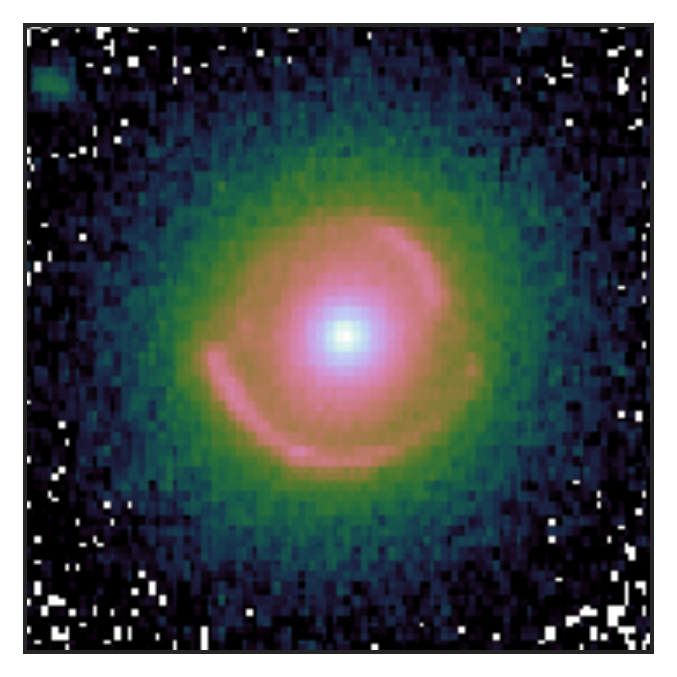}
		\end{minipage}	
		\begin{minipage}{.6\columnwidth}
            
        \end{minipage}
        \label{label}
	\end{subfigure}
	
	\begin{subfigure}[m]{\columnwidth}
		\begin{minipage}{.35\columnwidth}
			\includegraphics[width=\columnwidth]{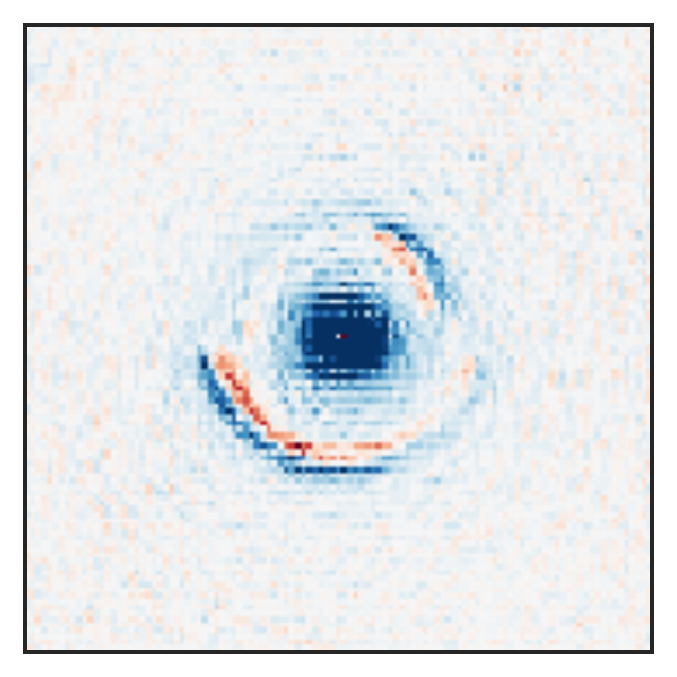}	
		\end{minipage}
		\begin{minipage}{.6\columnwidth}
            \caption{Take pixel-wise radial gradient outward from the centre of the deflector. The pixels with positive gradient (red) mark the inner edge of the arcs.}
        \end{minipage}
        \label{label}
	\end{subfigure}
	
	\begin{subfigure}[m]{\columnwidth}
		\begin{minipage}{.35\columnwidth}
			\includegraphics[width=\columnwidth]{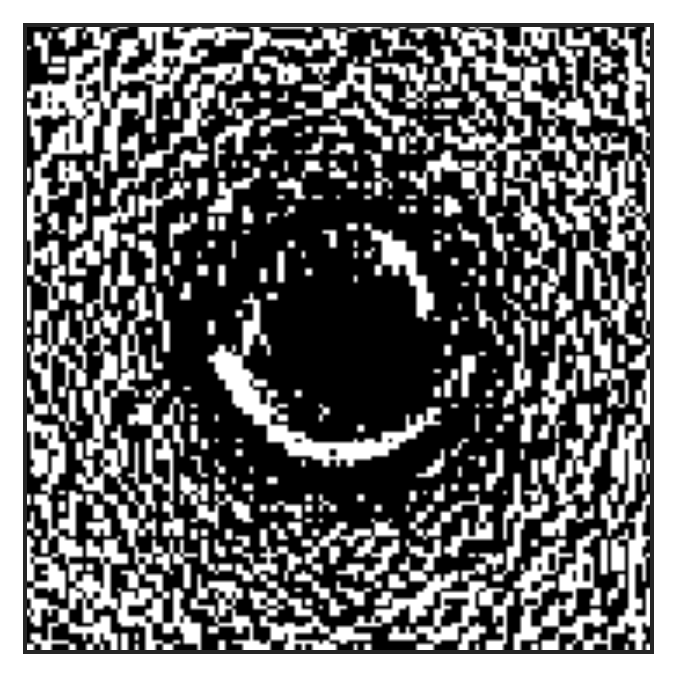}	
		\end{minipage}
		\begin{minipage}{.6\columnwidth}
            \caption{Make a binary image of the gradient with positive values set to 1 (white) and negative values set to 0 (black). Set all the pixels within the inner 0.4 arcsec to 0.}
        \end{minipage}
        \label{label}
	\end{subfigure}

	\begin{subfigure}[m]{\columnwidth}
		\begin{minipage}{.35\columnwidth}
			\includegraphics[width=\columnwidth]{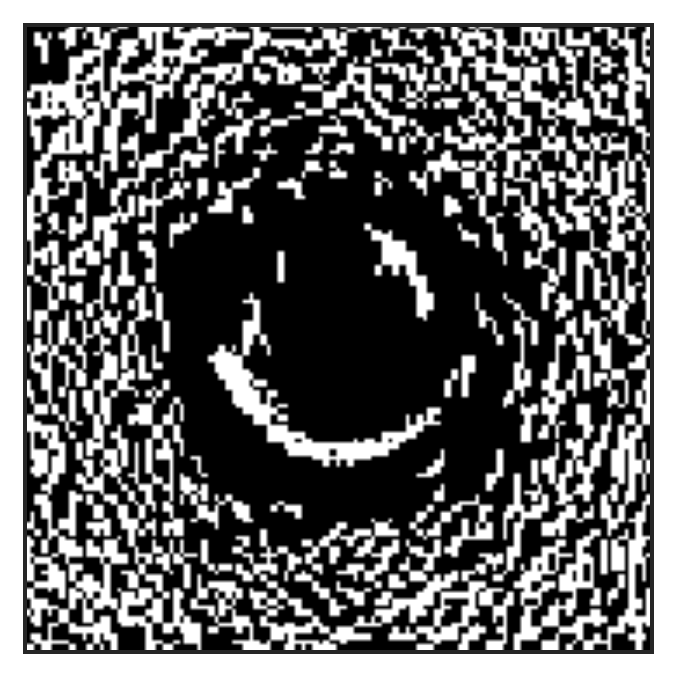}	
		\end{minipage}
		\begin{minipage}{.6\columnwidth}
            \caption{Set connected white regions with area below  5 pixels to zero. This procedure removes small white regions near the arc that were created due to noise.}
        \end{minipage}
        \label{label}
	\end{subfigure}

	\begin{subfigure}[m]{\columnwidth}
		\begin{minipage}{.35\columnwidth}
			\includegraphics[width=\columnwidth]{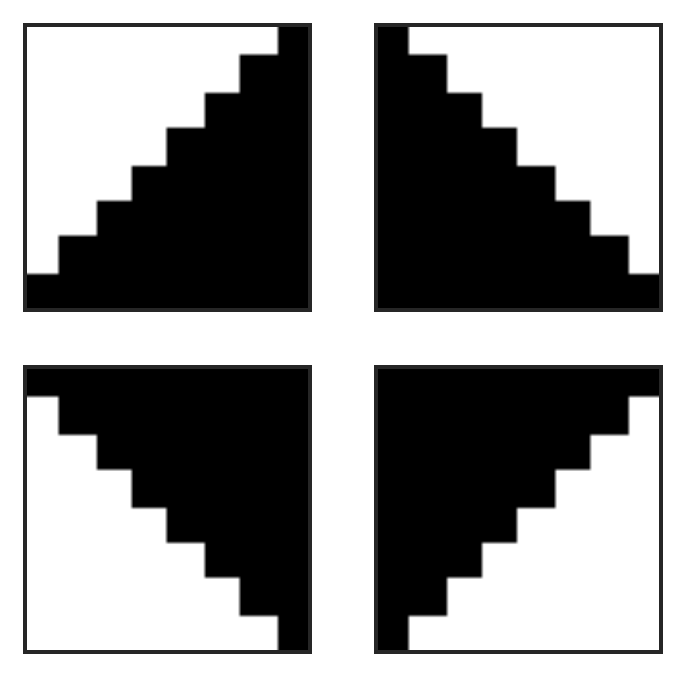}	
		\end{minipage}
		\begin{minipage}{.6\columnwidth}
            \caption{Dilate the binary image separately in the four quadrants with these structural elements. The size of the structural element matrices is 7$\times$7.}
        \end{minipage}
        \label{label}
	\end{subfigure}
	
	\begin{subfigure}[m]{\columnwidth}
		\begin{minipage}{.35\columnwidth}
			\includegraphics[width=\columnwidth]{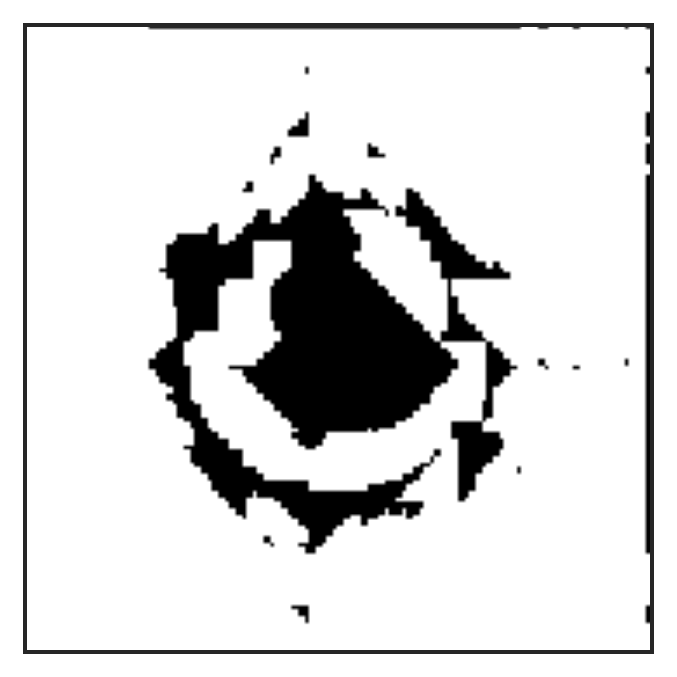}	
		\end{minipage}
		\begin{minipage}{.6\columnwidth}
            \caption{The dilation grows the white regions radially outward. Thus, the white regions can span over the whole width of the lensed arcs.}
        \end{minipage}
        \label{label}
	\end{subfigure}
	
	\begin{subfigure}[m]{\columnwidth}
		\begin{minipage}{.35\columnwidth}
			\includegraphics[width=\columnwidth]{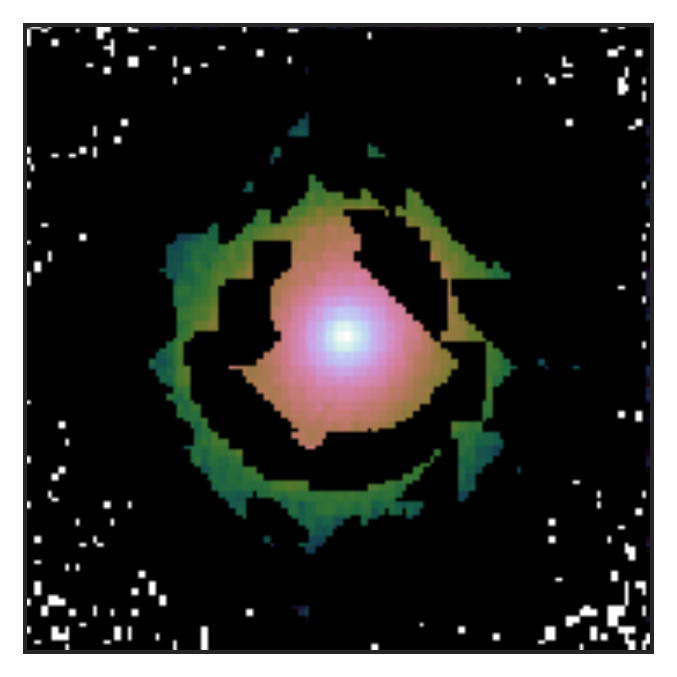}	
		\end{minipage}
		\begin{minipage}{.6\columnwidth}
            \caption{Invert the binary image to create the mask for the lensed arcs. The figure on the left shows the lens image with the lensed arcs masked.}
        \end{minipage}
        \label{label}
	\end{subfigure}
	
	\caption{ \label{fig:arc_mask}
	The algorithm to mask the lensed arcs. We use this mask to robustly fit the lens light profile only at one particular step within our model fitting procedure. The top figure shows an example lens to illustrate the procedure. All the subsequent figures illustrate different steps of the algorithm. The bottom figure shows the same lens image from the top figure with the lensed arcs masked. We used this mask to only fit the deflector's light distribution during an initial step of the optimization procedure.}
	
\end{figure}

We illustrate the algorithm for creating the mask for the lensed arcs in Figure \ref{fig:arc_mask}.

\renewcommand\thesubfigure{\arabic{subfigure}}

\section{Selected SLACS galaxies for modelling} \label{app:selected_galaxies}

We provide the list of selected galaxies in Table \ref{tab:sel_galaxies}.
\renewcommand{\arraystretch}{1.2}
\begin{table}
	\caption{\label{tab:sel_galaxies}
		\editth{List of SLACS lenses selected for modelling following the criteria described in Section \ref{sec:lens_sample}. The `Modelling success' column states whether we achieved a good quality model or not from our automated and uniform modelling approach.}
	}
	\begin{tabular}{lll}
	\hline
	Name & \textit{HST} camera and filter & Modelling success \\
	\hline	
	J0008$-$0004 & WFPC2 F606W & No \\
J0029$-$0055 & WFPC2 F606W & Yes \\
J0037$-$0942 & WFPC2 F606W & Yes \\
J0044$+$0113 & WFPC2 F606W & No \\
J0252$+$0039 & WFPC2 F606W & Yes \\
J0330$-$0020 & WFPC2 F606W & Yes \\
J0728$+$3835 & WFPC2 F606W & Yes \\
J0737$+$3216 & ACS F555W & Yes \\
J0819$+$4534 & WFPC2 F606W & No \\
J0903$+$4116 & WFPC2 F606W & Yes \\
J0912$+$0029 & ACS F555W & No \\
J0935$-$0003 & WFPC2 F606W & No \\
J0936$+$0913 & WFPC2 F606W & No \\
J0959$+$0410 & ACS F555W & Yes \\
J0959$+$4416 & WFPC2 F606W & No \\
J1016$+$3859 & WFPC2 F606W & No \\
J1020$+$1122 & WFPC2 F606W & No \\
J1023$+$4230 & WFPC2 F606W & No \\
J1100$+$5329 & WFPC2 F606W & No \\
J1112$+$0826 & WFPC2 F606W & Yes \\
J1134$+$6027 & WFPC2 F606W & No \\
J1142$+$1001 & WFPC2 F606W & No \\
J1143$-$0144 & ACS F555W & No \\
J1153$+$4612 & WFPC2 F606W & No \\
J1204$+$0358 & WFPC2 F606W & Yes \\
J1213$+$6708 & WFPC2 F606W & No \\
J1218$+$0830 & WFPC2 F606W & No \\
J1250$+$0523 & ACS F555W & Yes \\
J1306$+$0600 & WFPC2 F606W & Yes \\
J1313$+$4615 & WFPC2 F606W & Yes \\
J1319$+$1504 & WFPC2 F606W & No \\
J1402$+$6321 & ACS F555W & Yes \\
J1403$+$0006 & WFPC2 F606W & No \\
J1432$+$6317 & WFPC2 F606W & No \\
J1531$-$0105 & WFPC2 F606W & Yes \\
J1538$+$5817 & WFPC2 F606W & No \\
J1614$+$4522 & WFPC2 F606W & No \\
J1621$+$3931 & WFPC2 F606W & Yes \\
J1627$-$0053 & ACS F555W & Yes \\
J1630$+$4520 & ACS F555W & Yes \\
J1636$+$4707 & WFPC2 F606W & Yes \\
J1644$+$2625 & WFPC2 F606W & No \\
J2238$-$0754 & ACS F555W & Yes \\
J2300$+$0022 & ACS F555W & Yes \\
J2302$-$0840 & WFPC2 F606W & No \\
J2303$+$1422 & ACS F555W & Yes \\
J2321$-$0939 & WFPC2 F606W & No \\
J2341$+$0000 & WFPC2 F606W & No \\
J2343$-$0030 & WFPC2 F606W & Yes \\
J2347$-$0005 & WFPC2 F606W & No \\
	\hline
	\end{tabular}
\end{table}


\bsp	
\label{lastpage}
\end{document}